\documentclass[%
 reprint,
 amsmath,amssymb,
 aps,
]{revtex4-2}

\usepackage{graphicx}% Include figure files
\usepackage{dcolumn}% Align table columns on decimal point
\usepackage{bm}% bold math
\usepackage{xcolor}
\usepackage{xr}
\makeatletter

\newcommand*{\addFileDependency}[1]{% argument=file name and extension
\typeout{(#1)}% latexmk will find this if $recorder=0
\@addtofilelist{#1}
\IfFileExists{#1}{}{\typeout{No file #1.}}
}\makeatother

\newcommand*{\myexternaldocument}[1]{%
\externaldocument{#1}%
\addFileDependency{#1.tex}%
\addFileDependency{#1.aux}%
}
\myexternaldocument{supplementstuff}

\newcommand{\nuavg}[0]{\ensuremath{\left<\nu\right>}}
\newcommand{\nustar}[0]{\ensuremath{\nu^*}}
\newcommand{\pavg}[0]{\ensuremath{\left<p\right>}}
\newcommand{\pstar}[0]{\ensuremath{p^*}}

\newcommand{\qstruct}[0]{\ensuremath{q_8}}

\newcommand{\nsp}[0]{\ensuremath{n_\mathrm{sp}}}

\newcommand{\loss}[0]{\ensuremath{\mathcal{L}}}
\newcommand{\nbasins}[0]{\ensuremath{m}}

\begin{document}

\preprint{APS/123-QED}

\title{Designing athermal disordered solids with automatic differentiation}% Force line breaks with \\
%\thanks{Self-assembly of disordered solids}%

\author{Mengjie ZU}
 \altaffiliation{Institute of Science and Technology Austria}%Lines break automatically or can be forced with \\
 \email{mengjie.zu@ist.ac.at}
\author{Carl GOODRICH}%
 \email{carl.goodrich@ist.ac.at}
\affiliation{%
 Institute of Science and Technology Austria
}%

%\collaboration{MUSO Collaboration}%\noaffiliation

%\author{Charlie Author}
% \homepage{http://www.Second.institution.edu/~Charlie.Author}
%\affiliation{
% Second institution and/or address\\
% This line break forced% with \\
%}%
%\affiliation{
% Third institution, the second for Charlie Author
%}%
%\author{Delta Author}
%\affiliation{%
% Authors' institution and/or address\\
% This line break forced with \textbackslash\textbackslash
%}%

%\collaboration{CLEO Collaboration}%\noaffiliation

\date{\today}% It is always \today, today,
             %  but any date may be explicitly specified

\begin{abstract}
The ability to control forces between sub-micron-scale building blocks offers considerable potential for designing new materials through self-assembly. A typical paradigm is to first identify a particular (crystal) structure that has some desired property, and then design building-block interactions so that this structure assembles spontaneously. While significant theoretical and experimental progress has been made in assembling complicated structures in a variety of systems, this two-step paradigm fundamentally fails for structurally disordered solids, which lack a well-defined structure to use as a target. Here we show that disordered solids can still be treated from an inverse self-assembly perspective by targeting material properties directly. Using the Poisson's ratio, $\nu$, as a primary example, we show how differentiable programming connects experimentally relevant interaction parameters with emergent behavior, allowing us to iteratively ``train" the system until we find the set of interactions that leads to the Poisson's ratio we desire. Beyond the Poisson's ratio, we also tune the pressure and a measure of local 8-fold structural order, as well as multiple properties simultaneously, demonstrating the potential for nontrivial design in disordered solids. This approach is highly robust, transferable, and scalable, can handle a wide variety of model systems, properties of interest, and preparation dynamics, and can optimize over 100s or even 1000s of parameters. This result connects the fields of disordered solids and inverse self-assembly, indicating that many of the tools and ideas that have been developed to understand the assembly of crystals can also be used to control the properties of disordered solids. 
\begin{description}
\item[keywords]
self-assembly $|$ disordered solids $|$ auxetic materials $|$ simulation $|$ automatic differentiation
\end{description}
\end{abstract}

\maketitle

%\tableofcontents

\section{\label{sec:introduction} Introduction}

\begin{figure*}
\centering
\includegraphics[width=0.9\linewidth]{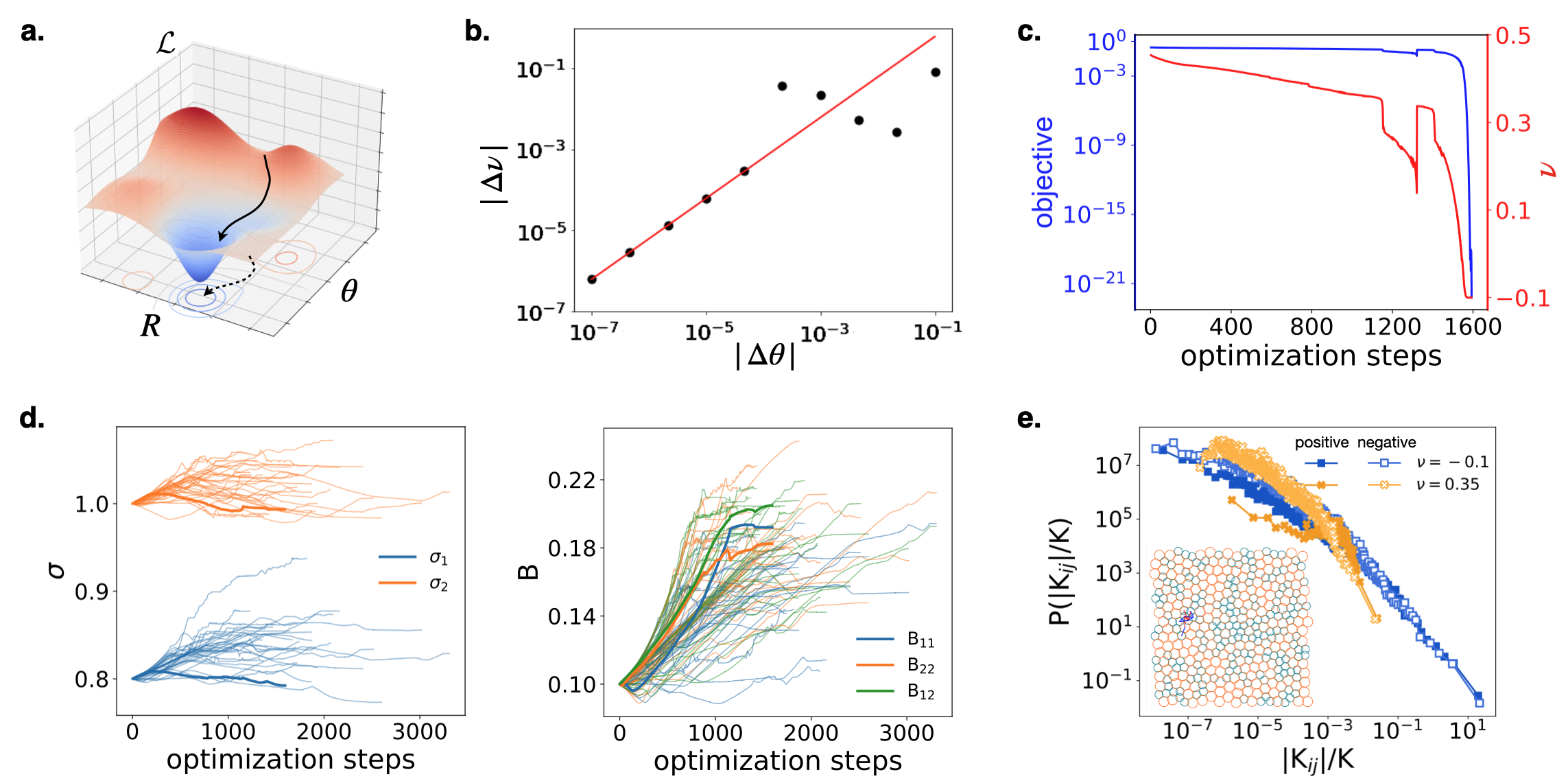}
\caption{\label{fig:Fig1} Tuning the Poisson's ratio of a single configuration. {\bf{a.}} Schematic of the landscape of the objective function $\loss$ in configuration space $R$ and parameter space $\theta$. Colored regions with contour lines represent the basins of attraction of distinct minima. The black arrow represents the configuration during the optimization process. {\bf{b.}} The gradient $\nabla_\theta \nu_b$ accurately predicts changes in the Poisson's ratio due to small changes in the model parameters. The red line shows the prediction $\left |\nabla_\theta \nu_b \cdot \Delta \theta \right|$, where $\Delta \theta$ is the change in parameters in an arbitrarily chosen direction, and the points show the measured change in the Poisson's ratio $\left| \nu_b(\Delta \theta) - \nu_b(\Delta \theta=0) \right|$ measured at $\Delta \theta$. Data is shown for a bidisperse system of 368 particles and binding energies 0.1, but the result is generic. {\bf{c.}} The evolution of the objective $\loss=(\nu_b-\nu^*)^2$ and Poisson’s ratio $\nu$ for an example optimization with target $\nu^*=-0.1$. The optimization is successful and converges in slightly less than 1600 optimization steps. Note: the spike around 1300 optimization steps corresponds to a discontinuous change in the position of the local minimum due to a structural rearrangement. 
{\bf{d.}} The evolution of the parameters for 35 successful optimizations. The thick lines correspond to the example shown in {\bf{c}}. The number of optimization steps is variable because the process terminates after convergence.
{\bf{e.}} The distribution of how each pair of particles contributes to the bulk modulus after training with $\nustar = -0.1$ (blue) and $\nustar = 0.35$ (orange). When $\nustar < 0.2$ (see SI for more comprehensive data), the positive and negative tails of this distribution broaden significantly. The positive and negative values are shown separately to demonstrate that, in this long tail, $P(K_{ij}) \approx P(-K_{ij})$. The inset shows the packing after training with $\nustar=-0.1$ and highlights the 10 bonds with the largest $|K_{ij}|$, with blue (red) lines indicating positive (negative) $K_{ij}$. This shows that the broad tails are spatially localized. }
\end{figure*}

%The transition from the previous paragraph to this depends on the journal (i.e. is the previous paragraph an abstract or a "first paragraph").
%One of the most challenging tasks in material design is to generate materials with robust and precisely controlled structures and properties. For crystals, structure and properties are closely connected, and so it usually suffices to focus exclusively on the structure. For disordered solids, however, there is no well-defined structure, yet bulk material properties are still well defined and reproducible. How much control do we have, in practice, over the properties of such solids when it is difficult to even quantify, let alone manipulate, their structure? 

To what extent can the material properties of disordered solids be controlled? Recent results show that the properties of random spring networks close to the isostatic point can be radically and precisely tuned through slight adjustments to the network topology~\cite{10.1103/physrevlett.114.225501,10.1039/c7sm01727h, 10.1073/pnas.1612139114,10.1073/pnas.1717442115, 10.1073/pnas.1806790116}. More specifically, the Poisson's ratio, $\nu$, of a randomly generated network can be tuned to either the upper or lower bounds by removing only $\sim 1\%$ of the springs -- the choice of which springs to remove determines the final value of $\nu$~\cite{10.1103/physrevlett.114.225501}. This is possible because each spring's contribution to the bulk modulus is independent of its contribution to the shear modulus, meaning that their ratio, which determines $\nu$, can be tuned by the choice of removed springs.
However, in a real material, it is usually not possible to make precise, targeted alterations to structure. In fact, the notion of removing springs from a network simply does not translate to materials made up of particulate building blocks, ranging from atoms or molecules to larger objects like proteins or colloids. Unlike a spring network, one cannot simply remove the interaction between two particular particles, let alone do so in a scalable way. %Instead, we need an approach for changing the structure of a particle-based material in a way that is statistically similar to the precise changes in the spring network. 

Nevertheless, this paper proposes a strategy for tuning the properties of particulate-based disordered solids. This strategy is akin to an inverse self-assembly approach, except that interaction parameters are adjusted to tune properties rather than structure. The necessary connection between interaction parameters and emergent properties is made by exploiting a class of numerical techniques called Automatic Differentiation (AD), enabling the exploration of high-dimensional and complicated design spaces. Once this connection is made, the system is ``trained" similar to how one trains a neural network.

%Using the ubiquitous coarse-grained model of DNA-coated spherical colloids, this paper demonstrates that controlling the properties of athermal disordered solids can be tackled from the perspective of inverse, multi-component self-assembly. 
We demonstrate this strategy by considering the simple case of athermal sticky spheres. Specifically, we consider a two-dimensional system of $N$ particles divided evenly into $\nsp$ species, where the diameter of each species, $\sigma_\alpha$, and the binding energy between each pair of species, $B_{\alpha\beta}=B_{\beta\alpha}$, can be continuously varied. 
Motivated by DNA-coated colloids, the particles interact via a Morse potential with the short-ranged repulsive part replaced with a finite soft repulsion (see Methods) ~\cite{10.1073/pnas.1109853108,10.1038/s41467-022-29853-w}.
%This model, discussed in Methods, is based on studies of jamming of soft spheres, but with the addition of species-dependent short-ranged attractions. 
The system is prepared at zero temperature following the protocol developed in the study of the jamming of soft spheres~\cite{10.1103/physreve.68.011306, 10.1146/annurev-conmatphys-070909-104045}, where particles are placed randomly (corresponding to infinite temperature) and then quenched to the nearest local energy minimum.

%The final material properties, such as the Poisson's ratio, clearly depend on the $\nsp + \nsp(\nsp+1)/2$ values of $D_\alpha$ and $B_{\alpha\beta}$, but we are left with three critically important questions: 1) Can these parameters be adjusted in order to accurately and precisely control material properties? (For example, how close can $\nu$ be tuned to a particular target $\nustar$?) 2) Can multiple properties be controlled simultaneously? and 3) How far can these properties be push away from their typical, untrained values? (For example, what is the lowest value of $\nu$ we can obtain with this approach?) We will show that, even for our exceedingly simple model, the answers to the first and second questions are a definitive yes. The final question is quantitative by nature and thus depends on the details of the model, the number of species, and the property in question. 

The final material properties, such as the Poisson's ratio, clearly depend on the $\nsp + \nsp(\nsp+1)/2$ values of $\sigma_\alpha$ and $B_{\alpha\beta}$, but we are left with two critically important questions. First, can these parameters be adjusted in order to accurately and precisely control material properties? For example, how close can $\nu$ be tuned to a particular target $\nustar$? Second, can multiple properties be controlled simultaneously and independently, enabling highly nontrivial design? We will show that, even for our exceedingly simple model, the answers to both questions are a definitive yes. 

This result is obtained by directly connecting the $\nsp + \nsp(\nsp+1)/2$ model parameters to changes in material properties. More precisely, we construct an objective function $\loss$, e.g. $\loss = (\nu - \nustar)^2$, that indicates how far a property (e.g. $\nu$) is from a target (e.g. $\nustar$), and use Automatic Differentiation (AD)~\cite{baydin2018automatic,10.1038/323533a0,10.1145/355586.364791} to calculate the gradient $\nabla_\theta \loss$ of $\loss$ with respect to the parameters $\theta = \{\sigma_\alpha\} \cup \{B_{\alpha\beta}\}$. $\nabla_\theta \loss$ indicates how changes to the parameters affects the objective, and there are numerous gradient-descent-based algorithms~\cite{ruder2017overview} for using this information to minimize $\loss$, thus tuning the property of interest. See Methods for more details.

At zero temperature, the $N$ particles arrange themselves into one of many possible ``configurations," or local minima of the potential energy landscape. In this paper, we will manipulate material properties first on the level of individual configurations, meaning we will pick a random configuration $b$ and tune the parameters until, \textit{ e.g.}, $\nu_b = \nustar$. We find that this works surprisingly well all the way to the perfectly auxetic limit $\nustar \rightarrow -1$. Interestingly, the Poisson's ratio of multiple configurations can also be tuned precisely and simultaneously, although this becomes more challenging as more configurations are optimized at the same time. We then consider the ensemble level, and again find a surprising ability to manipulate the \textit{average} Poisson's ratio $\nuavg$ within the range $0.2 < \nuavg < 0.7$. Importantly, this approach is highly general, and we use it to tune other properties and even multiple properties simultaneously.

%Our results are obtained by directly connecting the $\nsp + \nsp(\nsp+1)/2$ model parameters to changes in material properties. More precisely, we construct an objective function $\loss$, e.g. $\loss = (\nu - \nustar)^2$, that indicates how far a property (e.g. $\nu$) is from a target, and use Automatic Differentiation (AD) to calculate the gradient $\nabla_\theta \loss$ of $\loss$ with respect to the parameters $\theta = \{D_\alpha\} \cup \{B_{\alpha\beta}\}$. $\nabla_\theta \loss$ indicates how changes to the parameters affects the objective, and there are numerous gradient-descent-based algorithms for using this information to minimize $\loss$, thus controlling the property of interest. See Methods for more details.

%Say somewhere that we are focusing on the Poisson's ratio... justify this but also say that these results are general.

\section{Results}
\subsection{\label{sec:individuals}Training individual configurations}
%We are ultimately interested in manipulating ensemble average properties of disordered solids. For our system, this corresponds to a weighted average over local energy minima. For a property $X$, we have $\left<X\right> = \sum_b w_b X_b$, where the sum is over all local minima, $b$, in the energy landscape, $X_b$ is the property measured at minimum $b$, and $w_b$ is proportional to the high-dimensional volume of the basin of attraction (see Fig.~\ref{fig:Fig1}a). However, to begin, we focus on a single local minimum 

To begin, we select a configuration $b$ by placing $N$ particles randomly in a 2d periodic box and minimizing the energy to the nearest minimum. We then define the objective $\loss = (\nu_b - \nustar)^2$, where $\nu_b$ is the measured Poisson's ratio and $\nustar$ is the desired target chosen from between $-1$ and $1$, which are the theoretical bounds for isotropic systems in two dimensions.
Figure~\ref{fig:Fig1}b confirms that our AD-based calculation of $\nabla_\theta \nu_b$ accurately predicts the change in $\nu_b$ over finite changes in the parameters $\theta$. This is shown for a particular representative example where $N=368$ particles are evenly divided into $\nsp=2$ species with initial diameters of $0.8$ and $1.0$, a number density of $\rho=1.6$, and with constant binding energies $B_{\alpha\beta}=0.1$. Before training, we measure $\nu_b\approx 0.453$. We train the system using standard gradient-descent-based algorithms (see Methods) to iteritively adjust $\theta$ before recalculating $\nu_b$ and $\nabla_\theta \nu_b$. At each step, we reminimize the energy with respect to the particle positions, allowing us to track the configuration as parameters change. Figure~\ref{fig:Fig1}c shows how $\loss$ and $\nu$ change during 1600 iterations. 
%s process for a system of 368 particles composed of only two species and a target of $\nustar=-0.1$. The initial diameters are 0.8 and 1.0, all initial binding energies are $B_{\alpha\beta}=0.1$ ({\color{blue}in units of the scale of the repulsive interaction [[I don't think this is currently correct, ... ???]]}) \MZcomment{the length unit is discussed in Methods}, and the initial value of the Poisson's ratio is 0.453. 
Despite only having 5 parameters, the Poisson's ratio is tuned exactly to the target, meaning that the Poisson's ratio at this configuration is successfully and accurately controlled.

%What is special about the final parameters that makes the Poisson's ratio exactly $-0.1$? Unfortunately, this is difficult to discern. Fig.~\ref{fig:Fig1}d shows how the parameters evolve during training for 35 successfully trained configurations. 
Figure~\ref{fig:Fig1}d shows how the parameters evolve during training for 35 successfully trained configurations. Clearly, the final parameters depend strongly on the particular configuration; while the systems remain bidisperse and thus disordered, there is no obvious trend in the relative particle diameters nor in the three binding energies. 

Interestingly, for $\nustar$ less than approximately 0.2 or greater than approximately 0.7, the final Poisson's ratio seems to be dominated by a small localized region. To see this, we calculate $K_{ij}$ and $G_{ij}$, which are the contributions of the interaction between particles $i$ and $j$ to the bulk and shear moduli, respectively, so that $K=\sum_{ij}K_{ij}$ and $G=\sum_{ij}G_{ij}$ ~\cite{10.1103/physrevlett.114.225501,10.1039/c7sm01727h}. We observe a typical probability distribution for $P(K_{ij})$ and $P(G_{ij})$ when $0.2 < \nu < 0.7$ (orange data in Fig.~\ref{fig:Fig1}e), with no noticeable spatial correlations in the largest positive and negative values. However, for $\nu < 0.2$, the positive and negative tails of $P(K_{ij})$ widen considerably (blue data in Fig.~\ref{fig:Fig1}e) but equally so they largely offset each other, and we find that the largest values are spatially localized (inset). We hypothesize that the broad tails in $P(K_{ij})$ enable more dramatic tuning of $K$, allowing $K/G$ to become small and thus $\nu$ to become negative. Conversely, for $\nu > 0.7$, the tails of $P(G_{ij})$ broaden, enabling $K/G$ to become large. See SI for more comprehensive data. While we do not fully understand the origin of these extreme and localized regions, we note that the range of $\nu$ where they occur coincides with our ability to train the \textit{ensemble average} Poisson's ratio, see below.

% {\color{green}Interestingly, the final Poisson's ratio seems to be dominated by a relatively small region within the material (see Fig.~\ref{fig:Fig1}e and Supplementary Material), suggesting that the parameters may be fine-tuned to a particular structural motif that happens to appear in the configuration. 

% Maybe comment on the localization having an onset above nu=0.7 and below nu=0.2 (if this is the case).
% }
% \MZcomment{The probability density distribution of elastic modules also indicates localization for extreme targets. Achieving a small Poisson's ratio leads to a significant broadening of the tails in P(K). Conversely, obtaining a large Poisson's ratio results in a significant broadening of the tails in P(G). The comprehensive understanding of localization phenomenon  is currently unclear and will be explored in future work.}

%The implications of this on the ensemble average and the large system limit will be discussed later. 

\begin{figure}
\centering
\includegraphics[width=0.9\linewidth]{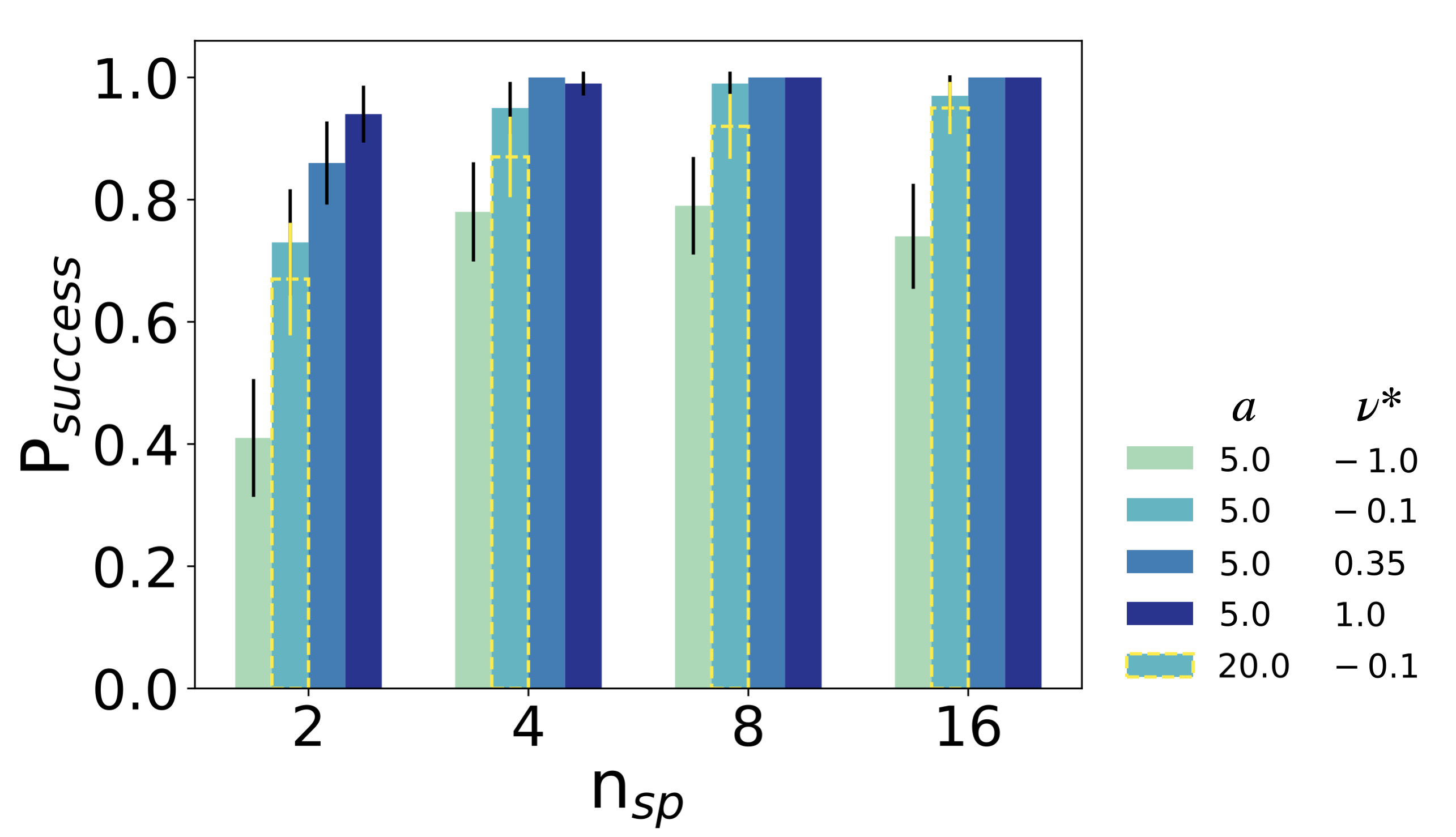}
\caption{\label{fig:Fig2} 
The ability to successfully train the Poisson's ratio at a given configuration depends on the number of species, \nsp, and the target Poisson's ratio, \nustar. The probability of successful training, $P_\mathrm{success}$, is close to 1 except for when $\nsp< 4$ or when $\nustar < -0.1$. Results are for a medium-ranged attractive potential ($a = 5$ in Eq.~\ref{eq:1}). Using a shorter-ranged potential ($a = 20$, yellow dashed line) decreases the ability to successfully train.
%\\\\
%{\color{red} The evaluation of DSPC optimisation for individual systems with various number of species, targets, and model parameter $\alpha$. {\bf{a.}} The yields of DSPCs  with different number of species $n_{sp}$ and target $\nu^*$ . The colours of the column refer to different targets, and the columns framed with yellow dashed line are results for systems with short-range length of attractive potential. 
%{\bf{b.}} The averaged validations of optimisation over 100 independent optimisation processes for systems with $n_{sp}=16$ types of particles, targeting the Poisson’s ratio in the range of theoretical values $[ -1.0, 1.0 ]$. The red cross corresponds to systems with initial parameters. 
%{\bf{c.}} The histogram of validation of 97 successful optimisation for systems with target $\nu^*=-0.1$ and $n_{sp}=16$, compared with the histogram of Poisson’s ratio of systems with initial parameters.
%}
}
\end{figure}

Figure~\ref{fig:Fig2} shows the probability $P_\mathrm{success}$ that we are able to successfully train a configuration of $N=368$ particles for various target Poisson's ratios and different numbers of species. Training is considered successful if the objective $\loss$ decreases below $\loss_\mathrm{thresh} = 10^{-6}$ within $10^4$ optimization steps. Not surprisingly, $P_\mathrm{success}$ increases with the number of species since that gives access to additional parameters, and with more conservative targets that are closer to the initial Poisson's ratio of approximately 0.5. We also find that the range of the attractive interaction matters, with shorter interactions making it slightly harder to train. 

\begin{figure}
\centering
\includegraphics[width=1.0\linewidth]{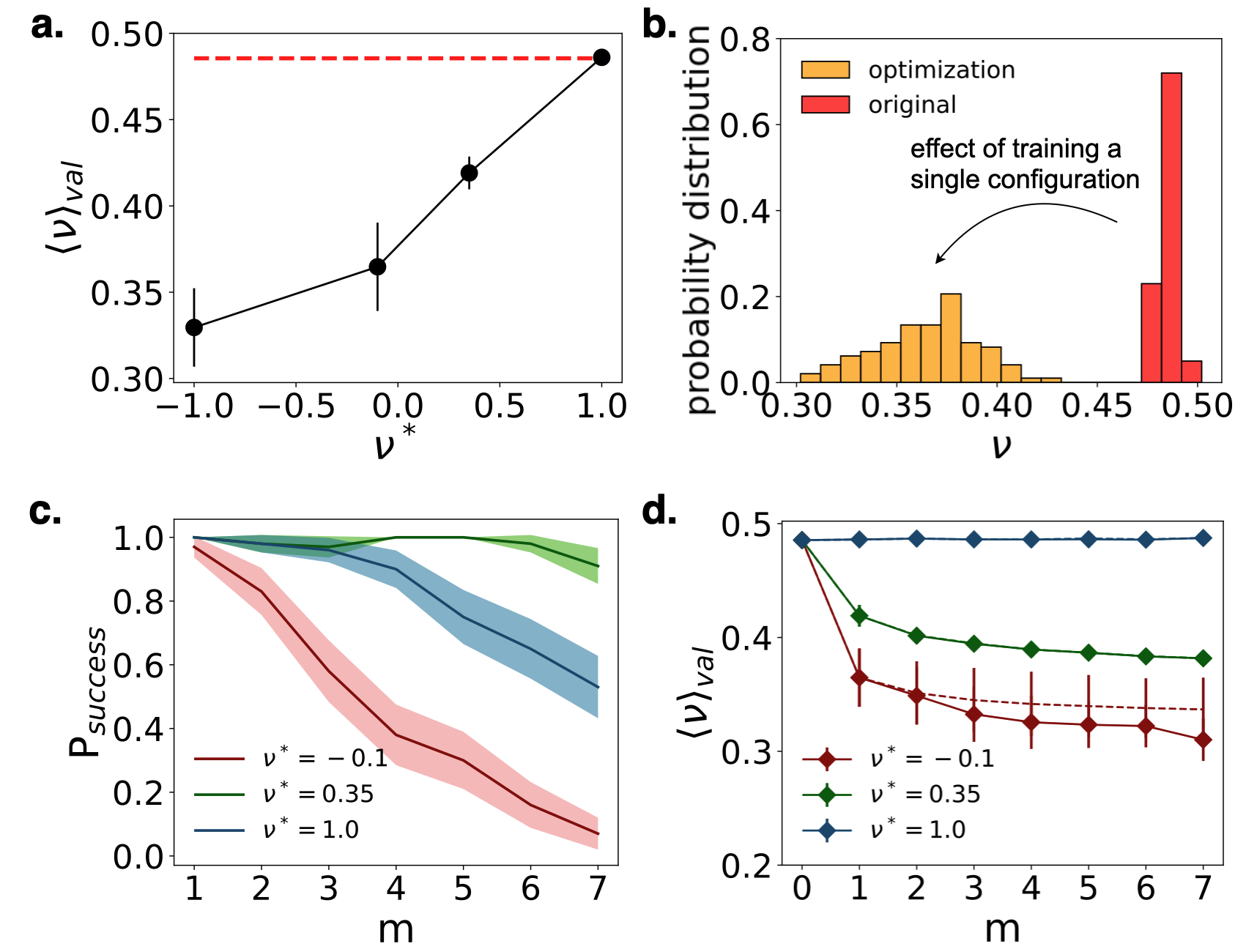}
\caption{\label{fig:Fig3} 
Training the Poisson's ratio at a single configuration, $\nu_b$, tunes the ensemble average Poisson's ratio, \nuavg. 
{\bf{a.}} $\nuavg$ averaged over 100 randomly chosen configurations using the final, learned parameters obtained from the $\nsp=16$ data from Fig.~\ref{fig:Fig2}. We call this the ``validation data" because, importantly, the Poisson's ratio from the configurations used for training were not included in these averages. The horizontal red line indicates the Poisson's ratio calculated with the initial parameters, so the difference between the points and the horizontal line indicates the amount of change in $\nuavg$.
{\bf{b.}} The full distribution of the validation data for the initial parameters (red) and the parameters obtained with $\nustar=-0.1$ (orange).
{\bf{c.}} $P_\mathrm{success}$ when targeting the Poisson's ratio of multiple configurations simultaneously. $P_\mathrm{success}$ decreases more rapidly for more aggressive $\nustar$. 
{\bf{d.}} Training multiple configurations increases the effect on $\nuavg$, though with diminishing returns above $\nbasins\approx 3$. $\nbasins=0$ means that the initial untrained parameters are used. The solid lines use only parameters after successful optimization, while the dashed lines use parameters after all optimizations. 
%\\\\
%{\color{red}The averaged validations of optimisation over 100 independent optimisation processes for systems with $n_{sp}=16$ types of particles, targeting the Poisson’s ratio in the range of theoretical values $[ -1.0, 1.0 ]$. The horizontal red line indicates the Poisson's ratio to systems with initial parameters. 
%The evaluation of DSPC optimisation for multiple basins for systems of $N=368$ particles grouped into $n_{sp}=16$ types of particles with model parameter $\alpha=5.0$.  
%{\bf{a.}} The changes of the yield of DSPCs with various number of basins and different target $\nu^*$. The yield is obtained by counting the number of successful optimizaion out of 100 independent optimisations.
%{\bf{b.}} The validations of the optimisation trained with different number of basins. The solid lines are the averaged validations of successful optimisation, the dashed lines are the averaged validations over all optimisations. Because of the high yields for target $\nu^*=0.35$ and $1.0$ the solid lines and dashed lines are almost overlapped.  
}
\end{figure}

These results show that we can accurately and precisely control $\nu_b$ over a wide range. Interestingly, Fig.~\ref{fig:Fig3}a-b show that the parameters learned through this process influence the entire ensemble of configurations even though they were never considered in the training process. For example, the parameters obtained from training a single system to $\nustar=-1$ results in $\nuavg = 0.33\pm 0.02$, well below the initial untrained value of $\nuavg =0.49$. This is a first indication that we can manipulate ensemble averages.

%While Figs.~\ref{fig:Fig1} and \ref{fig:Fig2} demonstrate the ability to accurately control the Poisson's ratio of a given configuration, we are ultimately interested in ensemble average properties. We have no reason to expect that any of the final parameters obtained so far should produce average Poisson's ratios even remotely close to the target because they were only trained on a single minimum. Nevertheless, Fig.~\ref{fig:Fig3}a-b show that the ensemble averages are still {\it influenced} by the training: for example, using the parameters obtained from training a single system to $\nustar=-1$ results in $\nuavg = 0.329\pm 0.022$, well below the initial untrained value of $\nuavg = 0.485$, albeit with a much larger variance. 

\begin{figure*}[htb]
\centering
\includegraphics[width=1.0\linewidth]{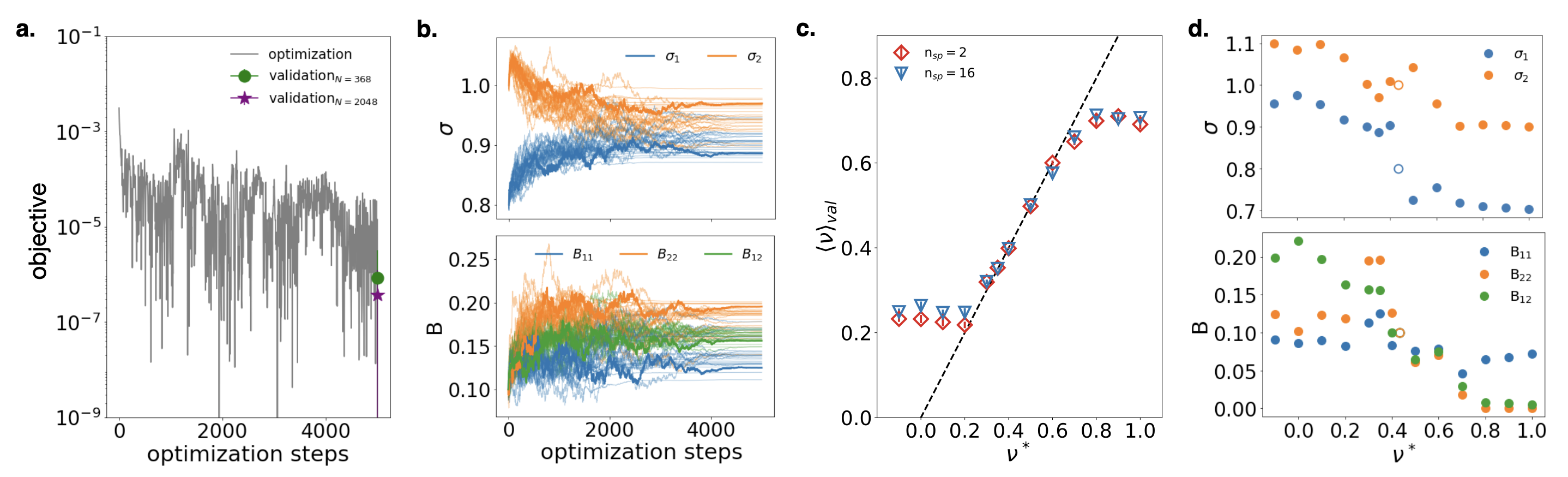}
\caption{\label{fig:Fig4} 
Tuning the ensemble.
%{\bf{a.}} Gradients of $\nuavg$ with respect to $\theta$ predict finite changes to $\nuavg$, even when the gradients are calculated using different configurations. The red line shows the prediction $\left |\nabla_\theta \nuavg \cdot \Delta \theta \right|$, with error bars, where $\Delta \theta$ is the change in parameters in an arbitrarily chosen direction, and the points show the measured change in the Poisson's ratio $\left| \nuavg(\Delta \theta) - \nuavg(\Delta \theta=0) \right|$ measured at $\Delta \theta$. For this plot only, we use $N=10000$ and average over $1000$ samples for each data point. 
{\bf{a.}} The objective (grey line) while optimizing $\nuavg$ with a target of $\nustar=0.35$ using $\nsp=2$ species. Only 8 systems are averaged at each step, making the objective very noisy. Nevertheless, proper validation averages of $\nuavg$ using the parameters obtained after 5000 optimization steps (green and purple points) show that the Poisson's ratio has indeed been tuned with very high accuracy. 
{\bf{b.}} The five parameters evolve in non-trivial ways over the course of optimization. The thick lines correspond to {\bf{a.}}, while the thin lines show the results of 29 other optimization runs with the same $\nustar$ that differ only in the random states that are sampled at each step. All 30 optimization attempts successfully lowered the objective below $10^{-5}$ and lead to similar, but not identical, trends in the final optimized parameters. 
{\bf{c.}} The final validated $\nuavg$ after optimization as a function of the target $\nustar$. We can accurately and precisely tune the Poisson's ratio over the range $0.2 < \nuavg < 0.7$. Interestingly, increasing the number of species does not improve our ability to train, suggesting that these limits might represent a fundamental barrier for sticky spheres. 
{\bf{d.}} The final parameters after optimization show clear trends with $\nustar$, indicating a general design strategy. 
%
%The green points show the validation data where 100 samples are averaged using the same parameters. The purple star gives the validated loss using larger systems ($N=2048$). The fact that the $N=2048$ validation loss is (reproducibly) lower than the $N=368$ validation loss indicates that the green data is dominated by fluctuations and that the system is even better tuned than this. 
%{\bf{c.}} The final validated loss forms a broad distribution that (marginally) improves with the number of species. [[Not really sure what to say here...]]
}
\end{figure*}

%Clearly, training on an individual minimum affects all minima. Perhaps unsurprisingly, 
This effect can be amplified by training on multiple configurations simultaneously. Specifically, we next choose a set of $\nbasins$ configurations, with $\nbasins$ ranging from 2 to 7, and define the objective $\loss = \sum_{b^\prime} ( \nu_{b^\prime} - \nustar)^2$, where here the sum is over the $\nbasins$ chosen configurations $b^\prime$. Figure~\ref{fig:Fig3}c shows that training becomes harder (less likely to succeed) as $\nbasins$ increases, especially for aggressive targets ($\nustar  <0$). Despite this, however, increasing $\nbasins$ does increase the effect on $\nuavg$, as shown by Fig.~\ref{fig:Fig3}d. %While this does not present a practical way to tune ensemble average properties, it further demonstrates the ability to manipulate ensemble averages. 

\subsection{Training ensemble-average quantities}
We now directly consider the ensemble average Poisson's ratio, $\nuavg$, by defining the objective $\loss = (\nuavg - \nustar)^2$. Note that this is not the same as the $\nbasins \rightarrow \infty$ limit of the objective from the previous section because here we are only concerned with the mean of the Poisson's ratio, not its value at every individual configuration. The other conceptual difference is that we can only ever estimate $\nuavg$ by averaging over a finite number of (randomly chosen) configurations, meaning that all calculations of $\loss$ and $\nabla_\theta \loss$ are necessarily stochastic. This is analogous to training in many machine learning models, where data is ``batched" and gradients are highly noisy. Importantly, as demonstrated in Fig.S1, average gradients of $\nuavg$ are predictive in the same way as in Fig.~\ref{fig:Fig1}b, allowing the use of stochastic gradient descent optimization.

Figure~\ref{fig:Fig4}a shows $\loss$ during optimization with a target of $\nustar=0.35$ using $\nsp=2$ species. At each step, we average over only 8 systems of size $N=368$, leading to large noise in $\loss$. This leads to a systematic overestimation of $\loss$ whenever the noise is larger than $\nuavg-\nustar$. Therefore, in order to ascertain how well the training has done, we also perform ``validation runs," where we use the final parameters obtained after 5000 optimization steps and calculate $\nuavg$ over a fresh set of 2000 configurations. As indicated by the green circle, $\loss_\mathrm{val} \approx 10^{-6}$, meaning that we have trained $\nuavg$ to $\nustar$ with an accuracy of 0.001. The purple star in Fig.~\ref{fig:Fig4}a shows an alternative validation where $\nuavg$ is calculated from 100 configurations of $N=2048$ particles each, demonstrating that even though our results are trained using small systems, the solution nevertheless applies to much larger systems. 

%Figure~S\ref{} shows that upon averaging, gradients of $\nuavg$ are predictive in the same way as in Fig.~\ref{fig:Fig1}b. Importantly, stochastic gradient descent algorithms are robust to overwhelming amounts of noise, and so in practice we average over only 10 systems of size $N=368$ at every gradient-descent iteration when training. This has two consequences. First, as shown in Fig.~\ref{fig:Fig4}a, our measured loss is very noisy during the optimization. In addition, the loss is also systematically inflated because noise in $\nuavg$ around $\nustar$ systematically increase $\loss=(\nuavg - \nustar)^2$. Thus, we also perform ``validation runs," where we use the final parameters obtained after 5000 optimization steps and calculate $\nuavg$ over a fresh set of XXX configurations (green circle). Finally, the purple star in Fig.~\ref{fig:Fig4}b shows an alternative validation where $\nuavg$ is calculated from 100 configurations of $N=2048$ particles each. 

Figure~\ref{fig:Fig4}b shows the 5 parameters at each optimization step (thick lines). The decrease in parameter fluctuations is caused by our variable learning rate (see SI), but the lack of systematic trends over the final 500 optimization steps suggests that the optimization has converged. 
%Despite the noise in $\loss$, Fig.~\ref{fig:Fig4}b suggests that the optimization has converged by showing the 5 parameters at each optimization step (thick lines). The decrease in parameter fluctuations is caused by our variable learning rate (see SI), but the con which stop fluctuating around 4500 steps. 
The thin lines show the same data for 29 other training runs that also target $\nustar=0.35$ and all achieve similarly small $\loss_\mathrm{val}$. Clearly the final parameters are not unique, meaning that the solution is degenerate, but there are still clear trends in the parameters (e.g. small particles bind more strongly to large particles than to other small particles). 

%Figure~\ref{fig:Fig4}X shows the final validated Poisson's ratio $\nuavg$ as a function of the target $\nustar$. 
Figure~\ref{fig:Fig4}c shows the final validated $\nuavg_\mathrm{val}$ as a function of the target $\nustar$, showing that we are able to successfully tune the Poisson's ratio between roughly 0.2 and 0.7. %Outside of this range, we are not able to get the desired Poisson's ratio. 
The existence of such bounds in our ability to tune $\nuavg$ is expected, especially for low $\nustar$ as creating low- and negative-Poisson's ratio materials is notoriously challenging~\cite{10.1103/physrevb.80.132104, PhysRevLett.101.085501}. We expect that more complicated systems, for example with non-spherical particles, could expand these bounds.
%and our sticky-sphere model is not equipped to generate the porous structures typically observed in auxetic materials. 
Notably, this range of roughly 0.2 to 0.7 coincides exactly with the onset of broad and localized bond-level response observed in Fig.~\ref{fig:Fig1}e. While training the Poisson's ratio of an individual configuration outside of this range is possible, it appears to rely on configuration-specific structural motifs that do not generalize to the full ensemble. This is also consistent with the observation in Fig.~\ref{fig:Fig1}d that the final parameters vary dramatically from one configuration to another. 

%: Mott and Roland~\cite{10.1103/physrevb.80.132104} argued that the observed dearth of isotropic materials with $\nu<0.2$~\footnote{This theoretical lower bound of 0.2 is for 3d materials, and it is unclear if our observed bound of 0.2 in 2 dimensions is a coincidence.} can be explained by considering roots of quadratic relations in classical elasticity.
%for a strict lower bound XXX \MZcomment{Roland 10.1103/PhysRevB.80.132104, please check if it is the paper you want to cite} suggests a lower bound to the Poisson's ratio of a system composed of spherically symmetric particles. 
%While we are able to engineer counter-examples to this when training individual configurations, this lower bound holds for ensemble averages as expected. 

Interestingly, we find that considering more species does not increase the range over which we can tune the system (blue triangles in Fig.~\ref{fig:Fig4}c). In principle, increasing the number of species should increase tunability, but there are no guarantees of this. It is unclear if improved training methodology could lead to better results with more species or if these bounds are imposed by physical constraints. Finally, Fig.~\ref{fig:Fig4}d shows the final parameters after 5000 optimization steps for different $\nustar$, suggesting a general design strategy for tuning $\nuavg$ over the range $0.2 < \nuavg < 0.7$. 

%What is the effect of including additional species? [[brief discussion.]]

%[[Brief discussion of the plot of $\left<p\right>_{val}$ vs $p^*$. Include any other quantities, maybe structural properties? Ok to include in supp material but mention here.]]

\begin{figure}
\centering
\includegraphics[width=0.9\linewidth]{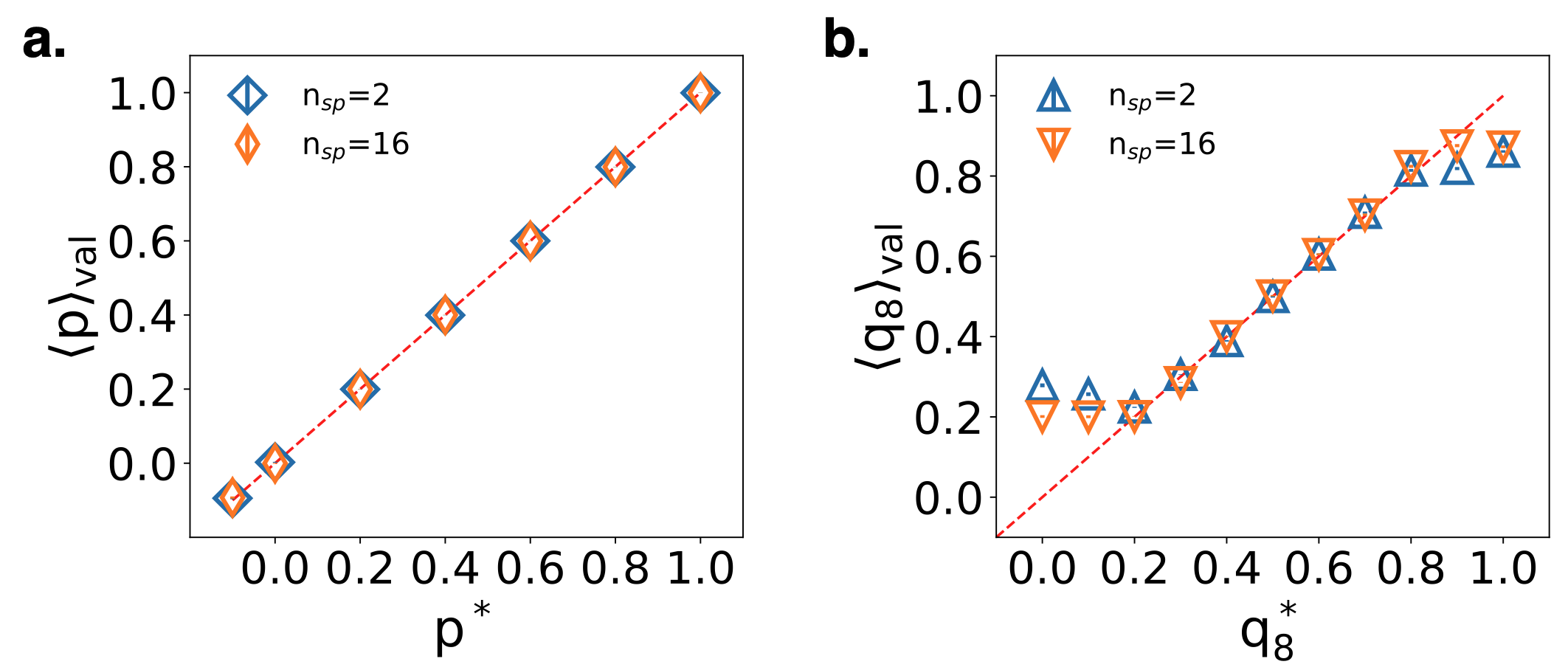}
\caption{\label{fig:Fig5} Tuning the ensemble-averaged pressure and a structural order parameter $\qstruct$. 
{\bf{a.}} {\bf{b.}} The validated $\left < p \right>$ and $\left< \qstruct \right>$ after optimization as functions of the target $\pstar$ and $q_8^*$ respectively. We demonstrate accurate and precise control of pressure throughout the test range, while fine-tuning $q_8$ within the range of $0.2 \leq \qstruct \leq 0.9$. %Due to finite size effects, the structural order parameter exceeds 0.0, and the quenching protocol prevents the formation of a perfect square lattice, resulting in a structural order parameter less than 1.0.
}
\end{figure}

\begin{figure}
\centering
\includegraphics[width=1.0\linewidth]{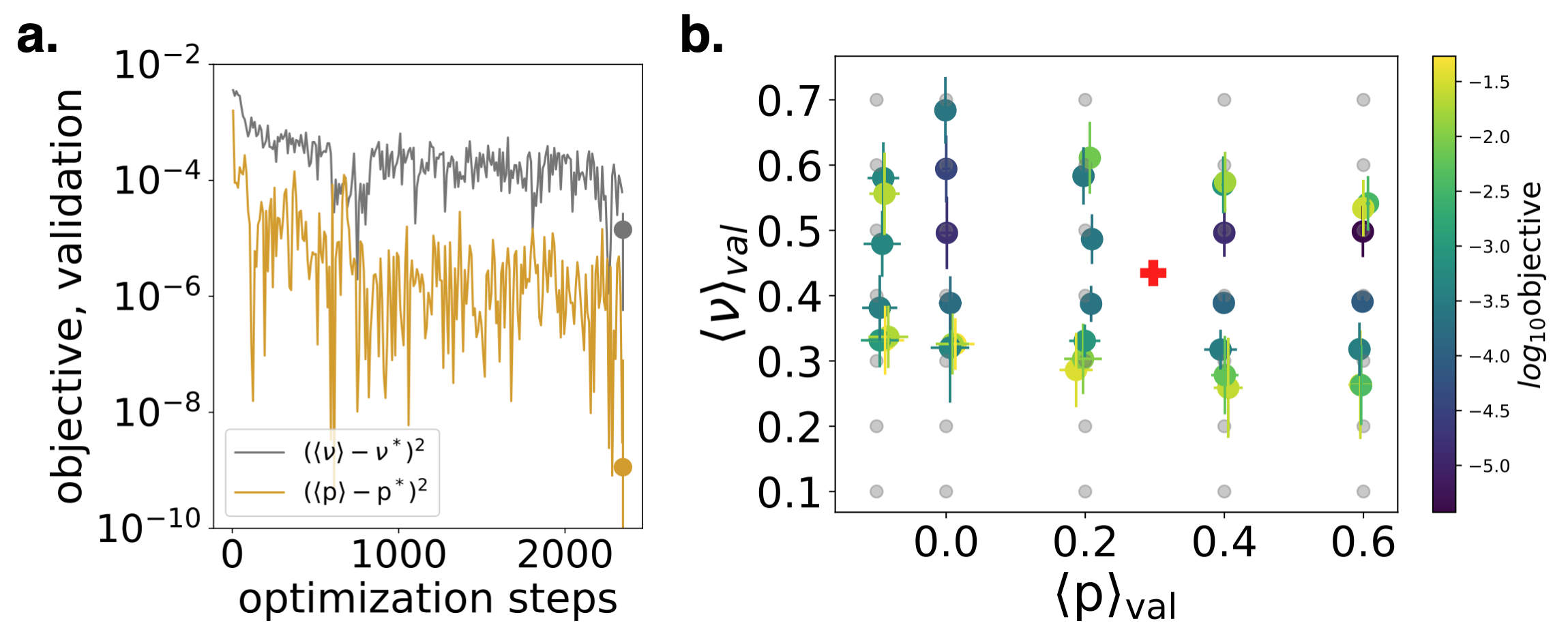}
\caption{\label{fig:Fig6} Tuning the ensemble-averaged Poisson’s ratio and pressure simultaneously. 
{\bf{a.}} The two components of the objective during optimization with $\nustar = 0.5$ and $\pstar=0.4$. 
{\bf{b.}} By changing $\nustar$ and $p^*$, we can tune the pair $(\nuavg,\pavg)$ over a non-trivial, 2 dimensional region. %This indicates a fundamental independence in the pressure and Poisson's ratio.
}
\end{figure}

%\begin{figure}
%\centering
%\includegraphics[width=1.0\linewidth]{Fig5.png}
%\caption{\label{fig:Fig5} The optimization for ensemble systems with tuning Poisson’s ratio and pressure simultaneously. 
%{\bf{a.}} The lines are the losses of an optimization process for ensemble systems, Poisson’s ratio and pressure respectively. The system consist s of N=368 particles evenly grouped into two species, interacting with Morse potential parameter $\alpha=6.0$. The optimisation process aim to minimise the loss function with target Poisson's ratio $\nu^*=0.5$ and target pressure $P^*=0.4$. The dots are validations, squared difference between targets and Poisson’s ratio (pressure) with optimised parameters at the final optimisation step, over 100 independent samples.
%{\bf{b.}} The validations for different sets of target. The red cross corresponds to the Poisson’s ratio and pressure with initial parameters. The validations are calculated with the parameters at the final optimisation steps, averaged over 100 samples of N=368 particles. The errorbars are the standard deviation of the validations. }
%\end{figure}

\subsection{Training multiple ensemble-average quantities simultaneously}
So far, we have focused on tuning the Poisson's ratio, but Fig.~\ref{fig:Fig5} shows that we can tune other quantities in exactly the same way, specifically the pressure and a structural order parameter $\qstruct$, which measures small amounts of local 8-fold symmetry in neighbor orientations (see SI). Now we ask whether we can tune multiple average properties {\it simultaneously}. Figure~\ref{fig:Fig6}a shows the simultaneous optimization of the Poisson's ratio ($\nustar=0.5$) and pressure ($\pstar=0.4$), which results in a final validated loss of $\loss_\mathrm{val} = 6\times 10^{-4}$, demonstrating success. To be precise, we use a combined objective function $\loss = (\nuavg - \nustar)^2 + (\pavg - \pstar)^2$, where the two components are plotted separately. 

Not all values of $(\pavg, \nuavg)$ can be obtained, and Fig.~\ref{fig:Fig6}b shows the final $(\pavg_\mathrm{val}, \nuavg_\mathrm{val})$ obtained from a series of systematic trials with different $(\pstar,\nustar)$ (indicated by light grey dots). This data shows a well-defined two-dimensional region of tunability, within which simultaneous optimization can be obtained. 

\section{Discussion}
We have shown that the properties of 2d athermal disordered solids can be inverse designed using Automatic Differentiation to connect properties of interest to particle-particle interactions. We have focused primarily on the Poisson's ratio, $\nu$, because as a unitless measure of elasticity it cannot be adjusted by trivially changing energy scales. Using a simple model of sticky spheres with tunable species-level interactions that is motivated by DNA-coated colloids~\cite{10.1073/pnas.1109853108,10.1038/s41467-022-29853-w}, we have shown that the Poisson's ratio of a single configuration can be tuned with a high success rate anywhere in the range $-1 \leq \nu \leq 1$ (the theoretical bounds for two-dimensional isotropic systems), and the ensemble average $\nuavg$ can be tuned in the range $0.2 < \nuavg < 0.7$. Furthermore, we can tune the pressure and local structural order in a similar way, and even design for multiple properties simultaneously. 

This can be compared with typical self-assembly paradigms, where one manipulates interactions so that, for example, a desired crystal structure becomes the thermodynamic ground state~\cite{10.1103/physrevlett.95.228301,10.1038/nmat4152,10.1073/pnas.1509316112,10.1021/acs.jpcb.8b05627,10.1063/1.5111492,10.1103/physrevlett.125.118003}. Instead, our approach targets the statistics of the metastable states. This has the distinct advantage that the assembly protocol (athermal relaxation in our case) is baked into the calculation of these statistics, and so we do not need to worry about kinetic traps or other barriers to assembly. 

While other systems exhibit a variable Poisson's ratio -- for example, jammed packings of soft spheres display $\nuavg \rightarrow 1$ near the unjamming transition~\footnote{Note that while our model reduces to the soft-sphere model by setting $B_{\alpha\beta}=0$ and $D_1 = 1.4 D_2$, the $\nuavg \rightarrow 1$ result is only obtained for finite systems in a fixed-pressure ensemble, and is therefore inaccessible in our fixed-volume ensemble.}~\cite{10.1146/annurev-conmatphys-070909-104045, 10.1088/0953-8984/22/3/033101}, our work demonstrates a new level of systematic and robust control. 
%This work is not the first to demonstrate a variable Poisson's ratio in disordered solids -- for example, jammed packings of soft spheres display $\nuavg \rightarrow 1$ near the unjamming transition~\footnote{Note that while our model reduces to the soft-sphere model by setting $B_{\alpha\beta}=0$ and $D_1 = 1.4 D_2$, the $\nuavg \rightarrow 1$ result is only obtained for finite systems in a fixed-pressure ensemble, and is therefore inaccessible in our fixed-volume ensemble.}~\cite{10.1146/annurev-conmatphys-070909-104045, 10.1088/0953-8984/22/3/033101}. 
In addition, we have shown that directly connecting objectives to parameters through the gradient $\nabla_\theta \loss$ enables the targeted design of \textit{multiple} properties, which is otherwise challenging. %Furthermore, gradient information can be used to \textit{predict} when various properties can be independently designed within a particular model. 
Together, this presents a scalable approach to inverse design and reveals a direct strategy for the targeted manipulation of disordered solids. 

Unlike for ensemble averages, we are able to precisely tune the Poisson's ratio of individual configurations with high success even in the auxetic or fully incompressible regimes. This is analogous to the bond-level tuning in Refs~\cite{10.1103/physrevlett.114.225501,10.1073/pnas.1717442115} except it is at the level of species-species interactions and maintains the constraint of force-balance on every particle. Figure~\ref{fig:Fig1}e also hints at an unexpected localized mechanism for creating auxetic materials. While these individual configurations are likely difficult to obtain experimentally due to the overwhelming number of competing structures, it is nevertheless telling how designable individual configurations are because it implies that the lower bound of $\nuavg$ is inherently a collective effect. At $\nuavg \approx 0.2$, any particular $\nu_b$ can still be lowered, but doing so necessarily increase the Poisson's ratio of other configurations so that $\nuavg$ is unchanged. 

One of the more surprising results is that increasing the number of species, and thus the number of parameters, does not increase the range over which we can tune $\nuavg$. This implies a fundamental constraint that is perhaps related to the general rarity of low and negative Poisson's ratio materials and leaves open the challenge of using our approach in more sophisticated models, with other types of design parameters, to obtain $\nuavg < 0$.

%\MZcomment{Our method efficiently produces auxetic and fully incompressible disordered solids, unconstrained by particle interaction potentials. This flexibility opens avenues for creating novel materials with unique attributes, such as anisotropic elasticity and high-energy absorption. What surprise is that our trained auxetic solids exhibit strong localized elastic responses, which provides a promising potential to generate general auxetic materials by training combined with local structures. The tuning of a few minimum simultaneously demonstrates the independent of effective dimension on the number of species, thus even with 2 species we could obtain well optimization results (see in SI). For ensemble training, our approach is analog to stochastic gradient descent, though the number of minimum is intractable with increasing system size, which can find general solutions. While the tuning mechanism behinds our approach is not fully understood, it holds significant promise by examining the interplay between structure, particle interactions, and properties of disordered solids that is a challenging and will  be explored in the future.}

This paper has only scratched the surface of how disordered solids can be designed with Automatic Differentiation. Exciting extensions of this include 1) considering more complicated building blocks, for example with nontrivial shapes or anisotropic potentials, %\MZcomment{how about 'considering more model parameters, such as building block shape, concentrations, and anisotropic potentials, etc.'}
2) considering different and more complicated preparation protocols, and 3) focusing on a wider range of material properties (e.g. nonlinear stress-strain behavior~\cite{10.1073/pnas.2119536119,10.1038/s41467-023-36965-4}, density of states, and allosteric responses~\cite{10.1073/pnas.1612139114}). Recent advances in differentiable programming ecosystems, including the Python packages JAX and JAX-MD used in this work, make it possible to incorporate Automatic Differentiation (AD) into many existing calculations. Important considerations, which are discussed in the context of the present work in the Methods section, include the predictiveness of gradients, strategies for avoiding large memory loads, and the importance of initial parameter guesses.

\bibliography{main}

%apsrev4-2.bst 2019-01-14 (MD) hand-edited version of apsrev4-1.bst
%Control: key (0)
%Control: author (8) initials jnrlst
%Control: editor formatted (1) identically to author
%Control: production of article title (0) allowed
%Control: page (0) single
%Control: year (1) truncated
%Control: production of eprint (0) enabled
\begin{thebibliography}{31}%
\makeatletter
\providecommand \@ifxundefined [1]{%
 \@ifx{#1\undefined}
}%
\providecommand \@ifnum [1]{%
 \ifnum #1\expandafter \@firstoftwo
 \else \expandafter \@secondoftwo
 \fi
}%
\providecommand \@ifx [1]{%
 \ifx #1\expandafter \@firstoftwo
 \else \expandafter \@secondoftwo
 \fi
}%
\providecommand \natexlab [1]{#1}%
\providecommand \enquote  [1]{``#1''}%
\providecommand \bibnamefont  [1]{#1}%
\providecommand \bibfnamefont [1]{#1}%
\providecommand \citenamefont [1]{#1}%
\providecommand \href@noop [0]{\@secondoftwo}%
\providecommand \href [0]{\begingroup \@sanitize@url \@href}%
\providecommand \@href[1]{\@@startlink{#1}\@@href}%
\providecommand \@@href[1]{\endgroup#1\@@endlink}%
\providecommand \@sanitize@url [0]{\catcode `\\12\catcode `\$12\catcode `\&12\catcode `\#12\catcode `\^12\catcode `\_12\catcode `\%12\relax}%
\providecommand \@@startlink[1]{}%
\providecommand \@@endlink[0]{}%
\providecommand \url  [0]{\begingroup\@sanitize@url \@url }%
\providecommand \@url [1]{\endgroup\@href {#1}{\urlprefix }}%
\providecommand \urlprefix  [0]{URL }%
\providecommand \Eprint [0]{\href }%
\providecommand \doibase [0]{https://doi.org/}%
\providecommand \selectlanguage [0]{\@gobble}%
\providecommand \bibinfo  [0]{\@secondoftwo}%
\providecommand \bibfield  [0]{\@secondoftwo}%
\providecommand \translation [1]{[#1]}%
\providecommand \BibitemOpen [0]{}%
\providecommand \bibitemStop [0]{}%
\providecommand \bibitemNoStop [0]{.\EOS\space}%
\providecommand \EOS [0]{\spacefactor3000\relax}%
\providecommand \BibitemShut  [1]{\csname bibitem#1\endcsname}%
\let\auto@bib@innerbib\@empty
%</preamble>
\bibitem [{\citenamefont {Goodrich}\ \emph {et~al.}(2015)\citenamefont {Goodrich}, \citenamefont {Liu},\ and\ \citenamefont {Nagel}}]{10.1103/physrevlett.114.225501}%
  \BibitemOpen
  \bibfield  {author} {\bibinfo {author} {\bibfnamefont {C.~P.}\ \bibnamefont {Goodrich}}, \bibinfo {author} {\bibfnamefont {A.~J.}\ \bibnamefont {Liu}},\ and\ \bibinfo {author} {\bibfnamefont {S.~R.}\ \bibnamefont {Nagel}},\ }\bibfield  {title} {\bibinfo {title} {{The Principle of Independent Bond-Level Response: Tuning by Pruning to Exploit Disorder for Global Behavior}},\ }\href {https://doi.org/10.1103/physrevlett.114.225501} {\bibfield  {journal} {\bibinfo  {journal} {Physical Review Letters}\ }\textbf {\bibinfo {volume} {114}},\ \bibinfo {pages} {225501} (\bibinfo {year} {2015})}\BibitemShut {NoStop}%
\bibitem [{\citenamefont {Hexner}\ \emph {et~al.}(2017)\citenamefont {Hexner}, \citenamefont {Liu},\ and\ \citenamefont {Nagel}}]{10.1039/c7sm01727h}%
  \BibitemOpen
  \bibfield  {author} {\bibinfo {author} {\bibfnamefont {D.}~\bibnamefont {Hexner}}, \bibinfo {author} {\bibfnamefont {A.~J.}\ \bibnamefont {Liu}},\ and\ \bibinfo {author} {\bibfnamefont {S.~R.}\ \bibnamefont {Nagel}},\ }\bibfield  {title} {\bibinfo {title} {{Role of local response in manipulating the elastic properties of disordered solids by bond removal}},\ }\href {https://doi.org/10.1039/c7sm01727h} {\bibfield  {journal} {\bibinfo  {journal} {Soft Matter}\ }\textbf {\bibinfo {volume} {14}},\ \bibinfo {pages} {312} (\bibinfo {year} {2017})}\BibitemShut {NoStop}%
\bibitem [{\citenamefont {Rocks}\ \emph {et~al.}(2017)\citenamefont {Rocks}, \citenamefont {Pashine}, \citenamefont {Bischofberger}, \citenamefont {Goodrich}, \citenamefont {Liu},\ and\ \citenamefont {Nagel}}]{10.1073/pnas.1612139114}%
  \BibitemOpen
  \bibfield  {author} {\bibinfo {author} {\bibfnamefont {J.~W.}\ \bibnamefont {Rocks}}, \bibinfo {author} {\bibfnamefont {N.}~\bibnamefont {Pashine}}, \bibinfo {author} {\bibfnamefont {I.}~\bibnamefont {Bischofberger}}, \bibinfo {author} {\bibfnamefont {C.~P.}\ \bibnamefont {Goodrich}}, \bibinfo {author} {\bibfnamefont {A.~J.}\ \bibnamefont {Liu}},\ and\ \bibinfo {author} {\bibfnamefont {S.~R.}\ \bibnamefont {Nagel}},\ }\bibfield  {title} {\bibinfo {title} {{Designing allostery-inspired response in mechanical networks}},\ }\href {https://doi.org/10.1073/pnas.1612139114} {\bibfield  {journal} {\bibinfo  {journal} {Proceedings of the National Academy of Sciences}\ }\textbf {\bibinfo {volume} {114}},\ \bibinfo {pages} {2520} (\bibinfo {year} {2017})},\ \Eprint {https://arxiv.org/abs/1607.08562} {1607.08562} \BibitemShut {NoStop}%
\bibitem [{\citenamefont {Reid}\ \emph {et~al.}(2018)\citenamefont {Reid}, \citenamefont {Pashine}, \citenamefont {Wozniak}, \citenamefont {Jaeger}, \citenamefont {Liu}, \citenamefont {Nagel},\ and\ \citenamefont {Pablo}}]{10.1073/pnas.1717442115}%
  \BibitemOpen
  \bibfield  {author} {\bibinfo {author} {\bibfnamefont {D.~R.}\ \bibnamefont {Reid}}, \bibinfo {author} {\bibfnamefont {N.}~\bibnamefont {Pashine}}, \bibinfo {author} {\bibfnamefont {J.~M.}\ \bibnamefont {Wozniak}}, \bibinfo {author} {\bibfnamefont {H.~M.}\ \bibnamefont {Jaeger}}, \bibinfo {author} {\bibfnamefont {A.~J.}\ \bibnamefont {Liu}}, \bibinfo {author} {\bibfnamefont {S.~R.}\ \bibnamefont {Nagel}},\ and\ \bibinfo {author} {\bibfnamefont {J.~J.~d.}\ \bibnamefont {Pablo}},\ }\bibfield  {title} {\bibinfo {title} {{Auxetic metamaterials from disordered networks}},\ }\href {https://doi.org/10.1073/pnas.1717442115} {\bibfield  {journal} {\bibinfo  {journal} {Proceedings of the National Academy of Sciences}\ }\textbf {\bibinfo {volume} {115}},\ \bibinfo {pages} {E1384} (\bibinfo {year} {2018})},\ \Eprint {https://arxiv.org/abs/1710.02493} {1710.02493} \BibitemShut {NoStop}%
\bibitem [{\citenamefont {Rocks}\ \emph {et~al.}(2019)\citenamefont {Rocks}, \citenamefont {Ronellenfitsch}, \citenamefont {Liu}, \citenamefont {Nagel},\ and\ \citenamefont {Katifori}}]{10.1073/pnas.1806790116}%
  \BibitemOpen
  \bibfield  {author} {\bibinfo {author} {\bibfnamefont {J.~W.}\ \bibnamefont {Rocks}}, \bibinfo {author} {\bibfnamefont {H.}~\bibnamefont {Ronellenfitsch}}, \bibinfo {author} {\bibfnamefont {A.~J.}\ \bibnamefont {Liu}}, \bibinfo {author} {\bibfnamefont {S.~R.}\ \bibnamefont {Nagel}},\ and\ \bibinfo {author} {\bibfnamefont {E.}~\bibnamefont {Katifori}},\ }\bibfield  {title} {\bibinfo {title} {{Limits of multifunctionality in tunable networks}},\ }\href {https://doi.org/10.1073/pnas.1806790116} {\bibfield  {journal} {\bibinfo  {journal} {Proceedings of the National Academy of Sciences}\ }\textbf {\bibinfo {volume} {116}},\ \bibinfo {pages} {2506} (\bibinfo {year} {2019})},\ \Eprint {https://arxiv.org/abs/1805.00504} {1805.00504} \BibitemShut {NoStop}%
\bibitem [{\citenamefont {Rogers}\ and\ \citenamefont {Crocker}(2011)}]{10.1073/pnas.1109853108}%
  \BibitemOpen
  \bibfield  {author} {\bibinfo {author} {\bibfnamefont {W.~B.}\ \bibnamefont {Rogers}}\ and\ \bibinfo {author} {\bibfnamefont {J.~C.}\ \bibnamefont {Crocker}},\ }\bibfield  {title} {\bibinfo {title} {{Direct measurements of DNA-mediated colloidal interactions and their quantitative modeling}},\ }\href {https://doi.org/10.1073/pnas.1109853108} {\bibfield  {journal} {\bibinfo  {journal} {Proceedings of the National Academy of Sciences}\ }\textbf {\bibinfo {volume} {108}},\ \bibinfo {pages} {15687} (\bibinfo {year} {2011})}\BibitemShut {NoStop}%
\bibitem [{\citenamefont {Cui}\ \emph {et~al.}(2022)\citenamefont {Cui}, \citenamefont {Marbach}, \citenamefont {Zheng}, \citenamefont {Holmes-Cerfon},\ and\ \citenamefont {Pine}}]{10.1038/s41467-022-29853-w}%
  \BibitemOpen
  \bibfield  {author} {\bibinfo {author} {\bibfnamefont {F.}~\bibnamefont {Cui}}, \bibinfo {author} {\bibfnamefont {S.}~\bibnamefont {Marbach}}, \bibinfo {author} {\bibfnamefont {J.~A.}\ \bibnamefont {Zheng}}, \bibinfo {author} {\bibfnamefont {M.}~\bibnamefont {Holmes-Cerfon}},\ and\ \bibinfo {author} {\bibfnamefont {D.~J.}\ \bibnamefont {Pine}},\ }\bibfield  {title} {\bibinfo {title} {{Comprehensive view of microscopic interactions between DNA-coated colloids}},\ }\href {https://doi.org/10.1038/s41467-022-29853-w} {\bibfield  {journal} {\bibinfo  {journal} {Nature Communications}\ }\textbf {\bibinfo {volume} {13}},\ \bibinfo {pages} {2304} (\bibinfo {year} {2022})},\ \Eprint {https://arxiv.org/abs/2111.06468} {2111.06468} \BibitemShut {NoStop}%
\bibitem [{\citenamefont {O’Hern}\ \emph {et~al.}(2003)\citenamefont {O’Hern}, \citenamefont {Silbert}, \citenamefont {Liu},\ and\ \citenamefont {Nagel}}]{10.1103/physreve.68.011306}%
  \BibitemOpen
  \bibfield  {author} {\bibinfo {author} {\bibfnamefont {C.~S.}\ \bibnamefont {O’Hern}}, \bibinfo {author} {\bibfnamefont {L.~E.}\ \bibnamefont {Silbert}}, \bibinfo {author} {\bibfnamefont {A.~J.}\ \bibnamefont {Liu}},\ and\ \bibinfo {author} {\bibfnamefont {S.~R.}\ \bibnamefont {Nagel}},\ }\bibfield  {title} {\bibinfo {title} {{Jamming at zero temperature and zero applied stress: The epitome of disorder}},\ }\href {https://doi.org/10.1103/physreve.68.011306} {\bibfield  {journal} {\bibinfo  {journal} {Physical Review E}\ }\textbf {\bibinfo {volume} {68}},\ \bibinfo {pages} {011306} (\bibinfo {year} {2003})},\ \Eprint {https://arxiv.org/abs/cond-mat/0304421} {cond-mat/0304421} \BibitemShut {NoStop}%
\bibitem [{\citenamefont {Liu}\ and\ \citenamefont {Nagel}(2010)}]{10.1146/annurev-conmatphys-070909-104045}%
  \BibitemOpen
  \bibfield  {author} {\bibinfo {author} {\bibfnamefont {A.~J.}\ \bibnamefont {Liu}}\ and\ \bibinfo {author} {\bibfnamefont {S.~R.}\ \bibnamefont {Nagel}},\ }\bibfield  {title} {\bibinfo {title} {{The Jamming Transition and the Marginally Jammed Solid}},\ }\href {https://doi.org/10.1146/annurev-conmatphys-070909-104045} {\bibfield  {journal} {\bibinfo  {journal} {Annual Review of Condensed Matter Physics}\ }\textbf {\bibinfo {volume} {1}},\ \bibinfo {pages} {347} (\bibinfo {year} {2010})}\BibitemShut {NoStop}%
\bibitem [{\citenamefont {Baydin}\ \emph {et~al.}(2018)\citenamefont {Baydin}, \citenamefont {Pearlmutter}, \citenamefont {Radul},\ and\ \citenamefont {Siskind}}]{baydin2018automatic}%
  \BibitemOpen
  \bibfield  {author} {\bibinfo {author} {\bibfnamefont {A.~G.}\ \bibnamefont {Baydin}}, \bibinfo {author} {\bibfnamefont {B.~A.}\ \bibnamefont {Pearlmutter}}, \bibinfo {author} {\bibfnamefont {A.~A.}\ \bibnamefont {Radul}},\ and\ \bibinfo {author} {\bibfnamefont {J.~M.}\ \bibnamefont {Siskind}},\ }\href@noop {} {\bibinfo {title} {Automatic differentiation in machine learning: a survey}} (\bibinfo {year} {2018}),\ \Eprint {https://arxiv.org/abs/1502.05767} {arXiv:1502.05767 [cs.SC]} \BibitemShut {NoStop}%
\bibitem [{\citenamefont {Rumelhart}\ \emph {et~al.}(1986)\citenamefont {Rumelhart}, \citenamefont {Hinton},\ and\ \citenamefont {Williams}}]{10.1038/323533a0}%
  \BibitemOpen
  \bibfield  {author} {\bibinfo {author} {\bibfnamefont {D.~E.}\ \bibnamefont {Rumelhart}}, \bibinfo {author} {\bibfnamefont {G.~E.}\ \bibnamefont {Hinton}},\ and\ \bibinfo {author} {\bibfnamefont {R.~J.}\ \bibnamefont {Williams}},\ }\bibfield  {title} {\bibinfo {title} {{Learning representations by back-propagating errors}},\ }\href {https://doi.org/10.1038/323533a0} {\bibfield  {journal} {\bibinfo  {journal} {Nature}\ }\textbf {\bibinfo {volume} {323}},\ \bibinfo {pages} {533} (\bibinfo {year} {1986})}\BibitemShut {NoStop}%
\bibitem [{\citenamefont {Wengert}(1964)}]{10.1145/355586.364791}%
  \BibitemOpen
  \bibfield  {author} {\bibinfo {author} {\bibfnamefont {R.~E.}\ \bibnamefont {Wengert}},\ }\bibfield  {title} {\bibinfo {title} {A simple automatic derivative evaluation program},\ }\href {https://doi.org/10.1145/355586.364791} {\bibfield  {journal} {\bibinfo  {journal} {Commun. ACM}\ }\textbf {\bibinfo {volume} {7}},\ \bibinfo {pages} {463–464} (\bibinfo {year} {1964})}\BibitemShut {NoStop}%
\bibitem [{\citenamefont {Ruder}(2017)}]{ruder2017overview}%
  \BibitemOpen
  \bibfield  {author} {\bibinfo {author} {\bibfnamefont {S.}~\bibnamefont {Ruder}},\ }\href@noop {} {\bibinfo {title} {An overview of gradient descent optimization algorithms}} (\bibinfo {year} {2017}),\ \Eprint {https://arxiv.org/abs/1609.04747} {arXiv:1609.04747 [cs.LG]} \BibitemShut {NoStop}%
\bibitem [{\citenamefont {Mott}\ and\ \citenamefont {Roland}(2009)}]{10.1103/physrevb.80.132104}%
  \BibitemOpen
  \bibfield  {author} {\bibinfo {author} {\bibfnamefont {P.~H.}\ \bibnamefont {Mott}}\ and\ \bibinfo {author} {\bibfnamefont {C.~M.}\ \bibnamefont {Roland}},\ }\bibfield  {title} {\bibinfo {title} {{Limits to Poisson’s ratio in isotropic materials}},\ }\href {https://doi.org/10.1103/physrevb.80.132104} {\bibfield  {journal} {\bibinfo  {journal} {Physical Review B}\ }\textbf {\bibinfo {volume} {80}},\ \bibinfo {pages} {132104} (\bibinfo {year} {2009})},\ \Eprint {https://arxiv.org/abs/0909.4697} {0909.4697} \BibitemShut {NoStop}%
\bibitem [{\citenamefont {Rechtsman}\ \emph {et~al.}(2008)\citenamefont {Rechtsman}, \citenamefont {Stillinger},\ and\ \citenamefont {Torquato}}]{PhysRevLett.101.085501}%
  \BibitemOpen
  \bibfield  {author} {\bibinfo {author} {\bibfnamefont {M.~C.}\ \bibnamefont {Rechtsman}}, \bibinfo {author} {\bibfnamefont {F.~H.}\ \bibnamefont {Stillinger}},\ and\ \bibinfo {author} {\bibfnamefont {S.}~\bibnamefont {Torquato}},\ }\bibfield  {title} {\bibinfo {title} {Negative poisson's ratio materials via isotropic interactions},\ }\href {https://doi.org/10.1103/PhysRevLett.101.085501} {\bibfield  {journal} {\bibinfo  {journal} {Phys. Rev. Lett.}\ }\textbf {\bibinfo {volume} {101}},\ \bibinfo {pages} {085501} (\bibinfo {year} {2008})}\BibitemShut {NoStop}%
\bibitem [{\citenamefont {Rechtsman}\ \emph {et~al.}(2005)\citenamefont {Rechtsman}, \citenamefont {Stillinger},\ and\ \citenamefont {Torquato}}]{10.1103/physrevlett.95.228301}%
  \BibitemOpen
  \bibfield  {author} {\bibinfo {author} {\bibfnamefont {M.~C.}\ \bibnamefont {Rechtsman}}, \bibinfo {author} {\bibfnamefont {F.~H.}\ \bibnamefont {Stillinger}},\ and\ \bibinfo {author} {\bibfnamefont {S.}~\bibnamefont {Torquato}},\ }\bibfield  {title} {\bibinfo {title} {{Optimized Interactions for Targeted Self-Assembly: Application to a Honeycomb Lattice}},\ }\href {https://doi.org/10.1103/physrevlett.95.228301} {\bibfield  {journal} {\bibinfo  {journal} {Physical Review Letters}\ }\textbf {\bibinfo {volume} {95}},\ \bibinfo {pages} {228301} (\bibinfo {year} {2005})},\ \Eprint {https://arxiv.org/abs/cond-mat/0508495} {cond-mat/0508495} \BibitemShut {NoStop}%
\bibitem [{\citenamefont {Engel}\ \emph {et~al.}(2015)\citenamefont {Engel}, \citenamefont {Damasceno}, \citenamefont {Phillips},\ and\ \citenamefont {Glotzer}}]{10.1038/nmat4152}%
  \BibitemOpen
  \bibfield  {author} {\bibinfo {author} {\bibfnamefont {M.}~\bibnamefont {Engel}}, \bibinfo {author} {\bibfnamefont {P.~F.}\ \bibnamefont {Damasceno}}, \bibinfo {author} {\bibfnamefont {C.~L.}\ \bibnamefont {Phillips}},\ and\ \bibinfo {author} {\bibfnamefont {S.~C.}\ \bibnamefont {Glotzer}},\ }\bibfield  {title} {\bibinfo {title} {{Computational self-assembly of a one-component icosahedral quasicrystal}},\ }\href {https://doi.org/10.1038/nmat4152} {\bibfield  {journal} {\bibinfo  {journal} {Nature Materials}\ }\textbf {\bibinfo {volume} {14}},\ \bibinfo {pages} {109} (\bibinfo {year} {2015})}\BibitemShut {NoStop}%
\bibitem [{\citenamefont {Miskin}\ \emph {et~al.}(2016)\citenamefont {Miskin}, \citenamefont {Khaira}, \citenamefont {Pablo},\ and\ \citenamefont {Jaeger}}]{10.1073/pnas.1509316112}%
  \BibitemOpen
  \bibfield  {author} {\bibinfo {author} {\bibfnamefont {M.~Z.}\ \bibnamefont {Miskin}}, \bibinfo {author} {\bibfnamefont {G.}~\bibnamefont {Khaira}}, \bibinfo {author} {\bibfnamefont {J.~J.~d.}\ \bibnamefont {Pablo}},\ and\ \bibinfo {author} {\bibfnamefont {H.~M.}\ \bibnamefont {Jaeger}},\ }\bibfield  {title} {\bibinfo {title} {{Turning statistical physics models into materials design engines}},\ }\href {https://doi.org/10.1073/pnas.1509316112} {\bibfield  {journal} {\bibinfo  {journal} {Proceedings of the National Academy of Sciences}\ }\textbf {\bibinfo {volume} {113}},\ \bibinfo {pages} {34} (\bibinfo {year} {2016})},\ \Eprint {https://arxiv.org/abs/1510.05580} {1510.05580} \BibitemShut {NoStop}%
\bibitem [{\citenamefont {Chen}\ \emph {et~al.}(2018)\citenamefont {Chen}, \citenamefont {Zhang},\ and\ \citenamefont {Torquato}}]{10.1021/acs.jpcb.8b05627}%
  \BibitemOpen
  \bibfield  {author} {\bibinfo {author} {\bibfnamefont {D.}~\bibnamefont {Chen}}, \bibinfo {author} {\bibfnamefont {G.}~\bibnamefont {Zhang}},\ and\ \bibinfo {author} {\bibfnamefont {S.}~\bibnamefont {Torquato}},\ }\bibfield  {title} {\bibinfo {title} {{Inverse Design of Colloidal Crystals via Optimized Patchy Interactions}},\ }\href {https://doi.org/10.1021/acs.jpcb.8b05627} {\bibfield  {journal} {\bibinfo  {journal} {The Journal of Physical Chemistry B}\ }\textbf {\bibinfo {volume} {122}},\ \bibinfo {pages} {8462} (\bibinfo {year} {2018})}\BibitemShut {NoStop}%
\bibitem [{\citenamefont {Kumar}\ \emph {et~al.}(2019)\citenamefont {Kumar}, \citenamefont {Coli}, \citenamefont {Dijkstra},\ and\ \citenamefont {Sastry}}]{10.1063/1.5111492}%
  \BibitemOpen
  \bibfield  {author} {\bibinfo {author} {\bibfnamefont {R.}~\bibnamefont {Kumar}}, \bibinfo {author} {\bibfnamefont {G.~M.}\ \bibnamefont {Coli}}, \bibinfo {author} {\bibfnamefont {M.}~\bibnamefont {Dijkstra}},\ and\ \bibinfo {author} {\bibfnamefont {S.}~\bibnamefont {Sastry}},\ }\bibfield  {title} {\bibinfo {title} {{Inverse design of charged colloidal particle interactions for self assembly into specified crystal structures}},\ }\href {https://doi.org/10.1063/1.5111492} {\bibfield  {journal} {\bibinfo  {journal} {The Journal of Chemical Physics}\ }\textbf {\bibinfo {volume} {151}},\ \bibinfo {pages} {084109} (\bibinfo {year} {2019})},\ \Eprint {https://arxiv.org/abs/1905.11061} {1905.11061} \BibitemShut {NoStop}%
\bibitem [{\citenamefont {Romano}\ \emph {et~al.}(2020)\citenamefont {Romano}, \citenamefont {Russo}, \citenamefont {Kroc},\ and\ \citenamefont {Šulc}}]{10.1103/physrevlett.125.118003}%
  \BibitemOpen
  \bibfield  {author} {\bibinfo {author} {\bibfnamefont {F.}~\bibnamefont {Romano}}, \bibinfo {author} {\bibfnamefont {J.}~\bibnamefont {Russo}}, \bibinfo {author} {\bibfnamefont {L.}~\bibnamefont {Kroc}},\ and\ \bibinfo {author} {\bibfnamefont {P.}~\bibnamefont {Šulc}},\ }\bibfield  {title} {\bibinfo {title} {{Designing Patchy Interactions to Self-Assemble Arbitrary Structures}},\ }\href {https://doi.org/10.1103/physrevlett.125.118003} {\bibfield  {journal} {\bibinfo  {journal} {Physical Review Letters}\ }\textbf {\bibinfo {volume} {125}},\ \bibinfo {pages} {118003} (\bibinfo {year} {2020})},\ \Eprint {https://arxiv.org/abs/2007.15873} {2007.15873} \BibitemShut {NoStop}%
\bibitem [{Note1()}]{Note1}%
  \BibitemOpen
  \bibinfo {note} {Note that while our model reduces to the soft-sphere model by setting $B_{\alpha \beta }=0$ and $D_1 = 1.4 D_2$, the $\protect \ensuremath {\left <\nu \right >}\rightarrow 1$ result is only obtained for finite systems in a fixed-pressure ensemble, and is therefore inaccessible in our fixed-volume ensemble.}\BibitemShut {Stop}%
\bibitem [{\citenamefont {Hecke}(2010)}]{10.1088/0953-8984/22/3/033101}%
  \BibitemOpen
  \bibfield  {author} {\bibinfo {author} {\bibfnamefont {M.~v.}\ \bibnamefont {Hecke}},\ }\bibfield  {title} {\bibinfo {title} {{Jamming of soft particles: geometry, mechanics, scaling and isostaticity}},\ }\href {https://doi.org/10.1088/0953-8984/22/3/033101} {\bibfield  {journal} {\bibinfo  {journal} {Journal of Physics: Condensed Matter}\ }\textbf {\bibinfo {volume} {22}},\ \bibinfo {pages} {033101} (\bibinfo {year} {2010})},\ \Eprint {https://arxiv.org/abs/0911.1384} {0911.1384} \BibitemShut {NoStop}%
\bibitem [{\citenamefont {Wang}\ \emph {et~al.}(2022)\citenamefont {Wang}, \citenamefont {Zhang}, \citenamefont {Li}, \citenamefont {Gao},\ and\ \citenamefont {Li}}]{10.1073/pnas.2119536119}%
  \BibitemOpen
  \bibfield  {author} {\bibinfo {author} {\bibfnamefont {Y.}~\bibnamefont {Wang}}, \bibinfo {author} {\bibfnamefont {X.}~\bibnamefont {Zhang}}, \bibinfo {author} {\bibfnamefont {Z.}~\bibnamefont {Li}}, \bibinfo {author} {\bibfnamefont {H.}~\bibnamefont {Gao}},\ and\ \bibinfo {author} {\bibfnamefont {X.}~\bibnamefont {Li}},\ }\bibfield  {title} {\bibinfo {title} {{Achieving the theoretical limit of strength in shell-based carbon nanolattices}},\ }\href {https://doi.org/10.1073/pnas.2119536119} {\bibfield  {journal} {\bibinfo  {journal} {Proceedings of the National Academy of Sciences}\ }\textbf {\bibinfo {volume} {119}},\ \bibinfo {pages} {e2119536119} (\bibinfo {year} {2022})}\BibitemShut {NoStop}%
\bibitem [{\citenamefont {Cheng}\ \emph {et~al.}(2023)\citenamefont {Cheng}, \citenamefont {Zhu}, \citenamefont {Cheng}, \citenamefont {Cai}, \citenamefont {Liu}, \citenamefont {Yao}, \citenamefont {Zhang},\ and\ \citenamefont {Duan}}]{10.1038/s41467-023-36965-4}%
  \BibitemOpen
  \bibfield  {author} {\bibinfo {author} {\bibfnamefont {H.}~\bibnamefont {Cheng}}, \bibinfo {author} {\bibfnamefont {X.}~\bibnamefont {Zhu}}, \bibinfo {author} {\bibfnamefont {X.}~\bibnamefont {Cheng}}, \bibinfo {author} {\bibfnamefont {P.}~\bibnamefont {Cai}}, \bibinfo {author} {\bibfnamefont {J.}~\bibnamefont {Liu}}, \bibinfo {author} {\bibfnamefont {H.}~\bibnamefont {Yao}}, \bibinfo {author} {\bibfnamefont {L.}~\bibnamefont {Zhang}},\ and\ \bibinfo {author} {\bibfnamefont {J.}~\bibnamefont {Duan}},\ }\bibfield  {title} {\bibinfo {title} {{Mechanical metamaterials made of freestanding quasi-BCC nanolattices of gold and copper with ultra-high energy absorption capacity}},\ }\href {https://doi.org/10.1038/s41467-023-36965-4} {\bibfield  {journal} {\bibinfo  {journal} {Nature Communications}\ }\textbf {\bibinfo {volume} {14}},\ \bibinfo {pages} {1243} (\bibinfo {year} {2023})}\BibitemShut {NoStop}%
\bibitem [{\citenamefont {Bitzek}\ \emph {et~al.}(2006)\citenamefont {Bitzek}, \citenamefont {Koskinen}, \citenamefont {G\"ahler}, \citenamefont {Moseler},\ and\ \citenamefont {Gumbsch}}]{PhysRevLett.97.170201}%
  \BibitemOpen
  \bibfield  {author} {\bibinfo {author} {\bibfnamefont {E.}~\bibnamefont {Bitzek}}, \bibinfo {author} {\bibfnamefont {P.}~\bibnamefont {Koskinen}}, \bibinfo {author} {\bibfnamefont {F.}~\bibnamefont {G\"ahler}}, \bibinfo {author} {\bibfnamefont {M.}~\bibnamefont {Moseler}},\ and\ \bibinfo {author} {\bibfnamefont {P.}~\bibnamefont {Gumbsch}},\ }\bibfield  {title} {\bibinfo {title} {Structural relaxation made simple},\ }\href {https://doi.org/10.1103/PhysRevLett.97.170201} {\bibfield  {journal} {\bibinfo  {journal} {Phys. Rev. Lett.}\ }\textbf {\bibinfo {volume} {97}},\ \bibinfo {pages} {170201} (\bibinfo {year} {2006})}\BibitemShut {NoStop}%
\bibitem [{\citenamefont {Guénolé}\ \emph {et~al.}(2020)\citenamefont {Guénolé}, \citenamefont {Nöhring}, \citenamefont {Vaid}, \citenamefont {Houllé}, \citenamefont {Xie}, \citenamefont {Prakash},\ and\ \citenamefont {Bitzek}}]{10.1016/j.commatsci.2020.109584}%
  \BibitemOpen
  \bibfield  {author} {\bibinfo {author} {\bibfnamefont {J.}~\bibnamefont {Guénolé}}, \bibinfo {author} {\bibfnamefont {W.~G.}\ \bibnamefont {Nöhring}}, \bibinfo {author} {\bibfnamefont {A.}~\bibnamefont {Vaid}}, \bibinfo {author} {\bibfnamefont {F.}~\bibnamefont {Houllé}}, \bibinfo {author} {\bibfnamefont {Z.}~\bibnamefont {Xie}}, \bibinfo {author} {\bibfnamefont {A.}~\bibnamefont {Prakash}},\ and\ \bibinfo {author} {\bibfnamefont {E.}~\bibnamefont {Bitzek}},\ }\bibfield  {title} {\bibinfo {title} {{Assessment and optimization of the fast inertial relaxation engine (fire) for energy minimization in atomistic simulations and its implementation in lammps}},\ }\href {https://doi.org/10.1016/j.commatsci.2020.109584} {\bibfield  {journal} {\bibinfo  {journal} {Computational Materials Science}\ }\textbf {\bibinfo {volume} {175}},\ \bibinfo {pages} {109584} (\bibinfo {year} {2020})},\ \Eprint {https://arxiv.org/abs/1908.02038} {1908.02038} \BibitemShut {NoStop}%
\bibitem [{\citenamefont {Goodrich}\ \emph {et~al.}(2014)\citenamefont {Goodrich}, \citenamefont {Dagois-Bohy}, \citenamefont {Tighe}, \citenamefont {van Hecke}, \citenamefont {Liu},\ and\ \citenamefont {Nagel}}]{PhysRevE.90.022138}%
  \BibitemOpen
  \bibfield  {author} {\bibinfo {author} {\bibfnamefont {C.~P.}\ \bibnamefont {Goodrich}}, \bibinfo {author} {\bibfnamefont {S.}~\bibnamefont {Dagois-Bohy}}, \bibinfo {author} {\bibfnamefont {B.~P.}\ \bibnamefont {Tighe}}, \bibinfo {author} {\bibfnamefont {M.}~\bibnamefont {van Hecke}}, \bibinfo {author} {\bibfnamefont {A.~J.}\ \bibnamefont {Liu}},\ and\ \bibinfo {author} {\bibfnamefont {S.~R.}\ \bibnamefont {Nagel}},\ }\bibfield  {title} {\bibinfo {title} {Jamming in finite systems: Stability, anisotropy, fluctuations, and scaling},\ }\href {https://doi.org/10.1103/PhysRevE.90.022138} {\bibfield  {journal} {\bibinfo  {journal} {Phys. Rev. E}\ }\textbf {\bibinfo {volume} {90}},\ \bibinfo {pages} {022138} (\bibinfo {year} {2014})}\BibitemShut {NoStop}%
\bibitem [{\citenamefont {Schoenholz}\ and\ \citenamefont {Cubuk}(2020)}]{jaxmd2020}%
  \BibitemOpen
  \bibfield  {author} {\bibinfo {author} {\bibfnamefont {S.~S.}\ \bibnamefont {Schoenholz}}\ and\ \bibinfo {author} {\bibfnamefont {E.~D.}\ \bibnamefont {Cubuk}},\ }\bibfield  {title} {\bibinfo {title} {Jax m.d. a framework for differentiable physics},\ }in\ \href {https://papers.nips.cc/paper/2020/file/83d3d4b6c9579515e1679aca8cbc8033-Paper.pdf} {\emph {\bibinfo {booktitle} {Advances in Neural Information Processing Systems}}},\ Vol.~\bibinfo {volume} {33}\ (\bibinfo  {publisher} {Curran Associates, Inc.},\ \bibinfo {year} {2020})\BibitemShut {NoStop}%
\bibitem [{\citenamefont {Bradbury}\ \emph {et~al.}(2018)\citenamefont {Bradbury}, \citenamefont {Frostig}, \citenamefont {Hawkins}, \citenamefont {Johnson}, \citenamefont {Leary}, \citenamefont {Maclaurin}, \citenamefont {Necula}, \citenamefont {Paszke}, \citenamefont {Vander{P}las}, \citenamefont {Wanderman-{M}ilne},\ and\ \citenamefont {Zhang}}]{jax2018github}%
  \BibitemOpen
  \bibfield  {author} {\bibinfo {author} {\bibfnamefont {J.}~\bibnamefont {Bradbury}}, \bibinfo {author} {\bibfnamefont {R.}~\bibnamefont {Frostig}}, \bibinfo {author} {\bibfnamefont {P.}~\bibnamefont {Hawkins}}, \bibinfo {author} {\bibfnamefont {M.~J.}\ \bibnamefont {Johnson}}, \bibinfo {author} {\bibfnamefont {C.}~\bibnamefont {Leary}}, \bibinfo {author} {\bibfnamefont {D.}~\bibnamefont {Maclaurin}}, \bibinfo {author} {\bibfnamefont {G.}~\bibnamefont {Necula}}, \bibinfo {author} {\bibfnamefont {A.}~\bibnamefont {Paszke}}, \bibinfo {author} {\bibfnamefont {J.}~\bibnamefont {Vander{P}las}}, \bibinfo {author} {\bibfnamefont {S.}~\bibnamefont {Wanderman-{M}ilne}},\ and\ \bibinfo {author} {\bibfnamefont {Q.}~\bibnamefont {Zhang}},\ }\href {http://github.com/google/jax} {\bibinfo {title} {{JAX}: composable transformations of {P}ython+{N}um{P}y programs}} (\bibinfo {year} {2018})\BibitemShut {NoStop}%
\bibitem [{\citenamefont {Blondel}\ \emph {et~al.}(2021)\citenamefont {Blondel}, \citenamefont {Berthet}, \citenamefont {Cuturi}, \citenamefont {Frostig}, \citenamefont {Hoyer}, \citenamefont {Llinares-L{\'o}pez}, \citenamefont {Pedregosa},\ and\ \citenamefont {Vert}}]{jaxopt_implicit_diff}%
  \BibitemOpen
  \bibfield  {author} {\bibinfo {author} {\bibfnamefont {M.}~\bibnamefont {Blondel}}, \bibinfo {author} {\bibfnamefont {Q.}~\bibnamefont {Berthet}}, \bibinfo {author} {\bibfnamefont {M.}~\bibnamefont {Cuturi}}, \bibinfo {author} {\bibfnamefont {R.}~\bibnamefont {Frostig}}, \bibinfo {author} {\bibfnamefont {S.}~\bibnamefont {Hoyer}}, \bibinfo {author} {\bibfnamefont {F.}~\bibnamefont {Llinares-L{\'o}pez}}, \bibinfo {author} {\bibfnamefont {F.}~\bibnamefont {Pedregosa}},\ and\ \bibinfo {author} {\bibfnamefont {J.-P.}\ \bibnamefont {Vert}},\ }\bibfield  {title} {\bibinfo {title} {Efficient and modular implicit differentiation},\ }\href@noop {} {\bibfield  {journal} {\bibinfo  {journal} {arXiv preprint arXiv:2105.15183}\ } (\bibinfo {year} {2021})}\BibitemShut {NoStop}%
\end{thebibliography}%


%apsrev4-2.bst 2019-01-14 (MD) hand-edited version of apsrev4-1.bst
%Control: key (0)
%Control: author (8) initials jnrlst
%Control: editor formatted (1) identically to author
%Control: production of article title (0) allowed
%Control: page (0) single
%Control: year (1) truncated
%Control: production of eprint (0) enabled
\begin{thebibliography}{2}%
\makeatletter
\providecommand \@ifxundefined [1]{%
 \@ifx{#1\undefined}
}%
\providecommand \@ifnum [1]{%
 \ifnum #1\expandafter \@firstoftwo
 \else \expandafter \@secondoftwo
 \fi
}%
\providecommand \@ifx [1]{%
 \ifx #1\expandafter \@firstoftwo
 \else \expandafter \@secondoftwo
 \fi
}%
\providecommand \natexlab [1]{#1}%
\providecommand \enquote  [1]{``#1''}%
\providecommand \bibnamefont  [1]{#1}%
\providecommand \bibfnamefont [1]{#1}%
\providecommand \citenamefont [1]{#1}%
\providecommand \href@noop [0]{\@secondoftwo}%
\providecommand \href [0]{\begingroup \@sanitize@url \@href}%
\providecommand \@href[1]{\@@startlink{#1}\@@href}%
\providecommand \@@href[1]{\endgroup#1\@@endlink}%
\providecommand \@sanitize@url [0]{\catcode `\\12\catcode `\$12\catcode `\&12\catcode `\#12\catcode `\^12\catcode `\_12\catcode `\%12\relax}%
\providecommand \@@startlink[1]{}%
\providecommand \@@endlink[0]{}%
\providecommand \url  [0]{\begingroup\@sanitize@url \@url }%
\providecommand \@url [1]{\endgroup\@href {#1}{\urlprefix }}%
\providecommand \urlprefix  [0]{URL }%
\providecommand \Eprint [0]{\href }%
\providecommand \doibase [0]{https://doi.org/}%
\providecommand \selectlanguage [0]{\@gobble}%
\providecommand \bibinfo  [0]{\@secondoftwo}%
\providecommand \bibfield  [0]{\@secondoftwo}%
\providecommand \translation [1]{[#1]}%
\providecommand \BibitemOpen [0]{}%
\providecommand \bibitemStop [0]{}%
\providecommand \bibitemNoStop [0]{.\EOS\space}%
\providecommand \EOS [0]{\spacefactor3000\relax}%
\providecommand \BibitemShut  [1]{\csname bibitem#1\endcsname}%
\let\auto@bib@innerbib\@empty
%</preamble>
\bibitem [{\citenamefont {Goodrich}\ \emph {et~al.}(2014)\citenamefont {Goodrich}, \citenamefont {Dagois-Bohy}, \citenamefont {Tighe}, \citenamefont {van Hecke}, \citenamefont {Liu},\ and\ \citenamefont {Nagel}}]{PhysRevE.90.022138}%
  \BibitemOpen
  \bibfield  {author} {\bibinfo {author} {\bibfnamefont {C.~P.}\ \bibnamefont {Goodrich}}, \bibinfo {author} {\bibfnamefont {S.}~\bibnamefont {Dagois-Bohy}}, \bibinfo {author} {\bibfnamefont {B.~P.}\ \bibnamefont {Tighe}}, \bibinfo {author} {\bibfnamefont {M.}~\bibnamefont {van Hecke}}, \bibinfo {author} {\bibfnamefont {A.~J.}\ \bibnamefont {Liu}},\ and\ \bibinfo {author} {\bibfnamefont {S.~R.}\ \bibnamefont {Nagel}},\ }\bibfield  {title} {\bibinfo {title} {Jamming in finite systems: Stability, anisotropy, fluctuations, and scaling},\ }\href {https://doi.org/10.1103/PhysRevE.90.022138} {\bibfield  {journal} {\bibinfo  {journal} {Phys. Rev. E}\ }\textbf {\bibinfo {volume} {90}},\ \bibinfo {pages} {022138} (\bibinfo {year} {2014})}\BibitemShut {NoStop}%
\bibitem [{\citenamefont {Zu}\ \emph {et~al.}(2017)\citenamefont {Zu}, \citenamefont {Tan},\ and\ \citenamefont {Xu}}]{10.1038/s41467-017-02316-3}%
  \BibitemOpen
  \bibfield  {author} {\bibinfo {author} {\bibfnamefont {M.}~\bibnamefont {Zu}}, \bibinfo {author} {\bibfnamefont {P.}~\bibnamefont {Tan}},\ and\ \bibinfo {author} {\bibfnamefont {N.}~\bibnamefont {Xu}},\ }\bibfield  {title} {\bibinfo {title} {Forming quasicrystals by monodisperse soft core particles},\ }\href {https://doi.org/10.1038/s41467-017-02316-3} {\bibfield  {journal} {\bibinfo  {journal} {Nature Communications}\ }\textbf {\bibinfo {volume} {8}},\ \bibinfo {pages} {2089} (\bibinfo {year} {2017})}\BibitemShut {NoStop}%
\end{thebibliography}%

\clearpage

\section{\label{sec:Model}Model and Methods}

{\bf{Model.}}
We consider a system composed of $N$ particles divided evenly into $\nsp$ species. 
Particles interact via a ``harmonic-Morse" pairwise potential given by
\begin{equation}\label{eq:1}
    V(r)=
     \begin{cases}
      \frac{k}{2}(r-\bar \sigma)^2 - B, & r < \bar \sigma \\
    B(e^{-2a(r-\bar \sigma)}-2e^{-a(r-\bar \sigma)}), & r \geq \bar \sigma
    \end{cases}
\end{equation}
where $r$ is the center-center distance between two particles, $\bar \sigma$ is the mean of their diameters, $k$ characterizes the short-ranged repulsions, and $B$ determines the strength of the medium-range attractions, whose extent is proportional to $1/a$. Unless otherwise stated, we use $k=5.0$ and $a=5$ for all pairs of particles. However, $B$ and $\bar\sigma$ depend on the species type of the particles in question. Specifically, we independently vary the attractive strength $B_{\alpha\beta}$ for every pair of species $\alpha$ and $\beta$ (provided $B_{\alpha\beta}=B_{\beta\alpha}$), as well as the particle diameter $\sigma_\alpha$ (so that $\bar \sigma_{\alpha\beta} = (\sigma_\alpha + \sigma_\beta)/2$). We then use the XPLOR smoothing function to truncate $V(r)$ at a distance $r_\mathrm{cut} = \bar \sigma + 9.9/a$. This model is commonly used to describe, for example, DNA-coated colloids where diameters and binding affinities can be manipulated at the species level~\cite{10.1073/pnas.1109853108,10.1038/s41467-022-29853-w}. Note that we use a harmonic repulsive force to ensure numerical stability at a random set of initial position, and that the strength $k$ of this repulsion is decoupled from the attractive forces to allow $B_{\alpha\beta} \rightarrow 0$.

% We consider a system composed of $\nsp$ species $\mu_{\alpha}$, each with the same concentration $\eta_{\mu_{\alpha}}=1/n_{sp}$. The particles belonging to the species $\mu_{\alpha}$ share identical potential parameters and diameter. These particles interact via a pairwise potential, harmonic-Morse potential, as described below,
%  %which consists of a short-ranged attractive interaction modeled with a Morse potential, and a finite purely repulsion modeled with a harmonic potential
% \begin{equation}\label{eq:1}
%     V_{hmm}(r_{ij})=
%      \begin{cases}
%       \frac{k}{2}(r_{ij}-\sigma_{ij})^2 - B, & r_{ij} < \sigma_{ij} \\
%     B(e^{-2\alpha(r_{ij}-\sigma_{ij})}-2e^{-\alpha(r_{ij}-\sigma_{ij})}), & r_{ij} \geq \sigma_{ij}
%     \end{cases}
% \end{equation}
% where $r_{ij}$ is the distance between the centers of particles $i$ and $j$, and $\sigma_{ij}$ is the average of their diameters $\sigma_i$ and $\sigma_j$. The harmonic repulsive forces are characterized by the spring constant $k=5.0$, thus independent of attractive forces. %Morse potential is widely employed to represent the DNA-mediated short-ranged attractive interactions. 
% The attractive potential is determined by two model parameters: the binding energy $B_{ij}$ between particles $i$ and $j$ and $\alpha$, defining the strength and range of attractive potential. This model can be realized in experiments with DNA-coated colloids, allowing the adjustment of these two potential through temperature and colloid design~\cite{Rogers2011,Cui2022}. 

For a given set of parameters $\theta = \{\sigma_\alpha\} \cup \{B_{\alpha\beta}\}$, we obtain stable athermal structures by using the FIRE algorithm~\cite{PhysRevLett.97.170201,10.1016/j.commatsci.2020.109584} to minimize the total potential energy $E = \sum_{\left< ij \right>} V(r_{ij})$ starting from a random set of initial positions. We always consider an even number of species, and set the initial diameters for half the species to be $0.8$ times that of the other half, with an overall number density of $1.6$. The initial binding strengths are all set to $B_{\alpha\beta}=0.1$. 

%To generate compact disordered solids, a system of $N=368$ particles in a $50:50$ bidisperse mixture with sizes $0.8\sigma$ and $1.0\sigma$, respectively, and a number density $\rho=N/V=1.6\sigma^2$ is utilized, where $\sigma$ represents a length unit. The attractive length $x_a$ is independent of particle size and is defined as $x_a=9.9\sigma/\alpha$. The XPLOR smoothing function is employed to ensure both the potential and the force smoothly approach 0 at $r_{cut}=x_a + \sigma_{ij}$. All elements of the interaction matrix are set as $B_{ij}=0.1$. 

{\bf{Optimization for individual configurations.}}
The original configurations are generated by quenching random packings to local energy minima using FIRE algorithm. Once an initial energy-minimized state with original setting are reached, the Poisson's ratio $\nu$ is calculated explicitly using linear response~\cite{PhysRevE.90.022138}, and combined with the target Poisson's ratio $\nustar$ to construct the objective function $\loss=(\nu-\nustar)^2$.  

Our model, the energy minimization, and the calculation of $\nu$ and $\loss$ are performed entirely using the library JAXMD~\cite{jaxmd2020} and the Automatic Differentiation library JAX~\cite{jax2018github} on which it is built. As a result, the entire calculation is end-to-end differentiable, allowing us to accurately and efficiently calculate $\nabla_\theta \loss$. Implicit differentiation, implemented using the JAXOPT library~\cite{jaxopt_implicit_diff}, is used to propagate derivatives through the energy minimization as well as a key step in the linear response calculation, which ensures gradient accuracy and mitigates any extensive memory overhead. We then employ the RMSProp algorithm (implemented as part of JAX) to iteratively update the parameters $\theta$ based on the gradients. To improve the convergence speed, we dynamically adjust the learning rate using a meta-learning program that adaptively generates per-step hyperparameters. At each iteration, we use the final configuration of the previous step for the initial positions of the energy minimization process, ensuring that we consistently track a given energy minimum. Occasionally, the energy minimum transforms into a saddle point and the system undergoes a rearrangement. This results in a small spike in $\loss$, e.g. see Fig.~\ref{fig:Fig1}c at around 1300 steps.

%To obtain the gradients of the objective function, the energy minimization and the measurement of specific property are programmed into a differentiable process. This process is decorated for adding implicit differentiation using JAXOPT package[? ] to solve memory and chaos  issue induced by the large number of energy minization steps. In this work, we employ RMSProp[? ? ] with advantage of its fast convergence speed to minimize the objective function. The learning rate is another crucial factor for the optimization, we interpolate a meta-learning programme that can adaptively generate a per-step hyperparameter - learning rate to accelerate the convergence of the optimization. We repeat the measurement-optimization over a number of iterations until the objective function is less than a threshold value. At each iteration, the measurement is implemented after energy minimizing the previous packing with adjusted parameters.

{\bf{Optimization for ensemble-averaged quantities.}}
The primary difference in optimization between individual and ensemble systems lies in the input configurations. In the optimization process for ensemble-averaged quantities, we generate 8 new configurations (by minimizing from random initial positions) at every step. Given our relatively small system size, this is well below the number of configurations needed to obtain accurate ensemble averages, meaning our estimates of $\loss$ and $\nabla_\theta \loss$ are highly noisy. However, stochastic optimization algorithms like RMSProp are well-suited to very noisy gradients and this does not prevent convergence. 

\clearpage

\end{document}

\title{Supplementary information for ``Designing athermal disordered solids with automatic differentiation"}

\author{Mengjie ZU}
 \altaffiliation{Institute of Science and Technology Austria}%Lines break automatically or can be forced with \\
 \email{mengjie.zu@ist.ac.at}
\author{Carl GOODRICH}%
 \email{carl.goodrich@ist.ac.at}
\affiliation{%
 Institute of Science and Technology Austria
}%

\maketitle

\subsection{Gradient validation for the ensemble}
\begin{figure*}
\centering
\includegraphics[width=0.5\linewidth]{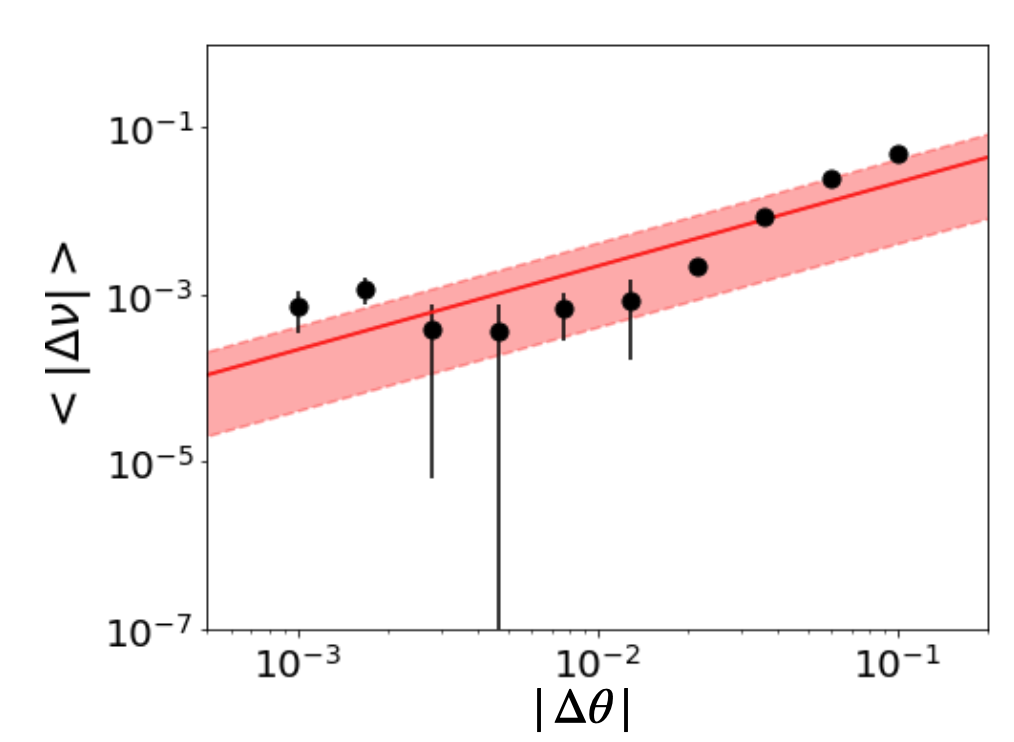}
\caption{ The averaged change in the Poisson's ratio $\left<|\Delta \nu|\right>$ as a function of parameter variations $\Delta \theta$ over 1000 configurations. The solid red line represents the averaged calculated gradient projected onto an arbitrarily chosen direction, with the shaded region indicating the estimated error.}
\label{fig:Figs1}
\end{figure*}

To demonstrate the predictiveness of the average gradient of the ensemble $\left<\nabla_{\theta}\nu\right>$, we measure the average change of the Poisson's ratio over 1000 configurations. The configurations are set up in the same way as in Fig 1b, but with a large number of particles ($N=10000$) to improve the statistics. In Fig.~\ref{fig:Figs1}, the dots show the averaged change in the Poisson's ratio $\left<|\Delta \nu|\right>=\left<|\nu(\Delta\theta)-\nu(\Delta\theta=0)|\right>$ measured at $\Delta \theta$,  where $\Delta \theta$ is the change in parameters in an arbitrarily chosen direction, and the red line shows the prediction $| \left<\nabla_{\theta}\nu\right>\cdot \Delta \theta|$. Due to the finite system size, the distribution of Poisson's ratio has a width with an order of magnitude of $-4$. Therefore, only results above $10^{-3}$ were displayed. Despite the fluctuations, the averaged calculated gradient was still able to predict the change in the Poisson's ratio for the ensemble.

\subsection{Number of optimization steps}

\begin{figure}
\centering
\includegraphics[width=0.5\linewidth]{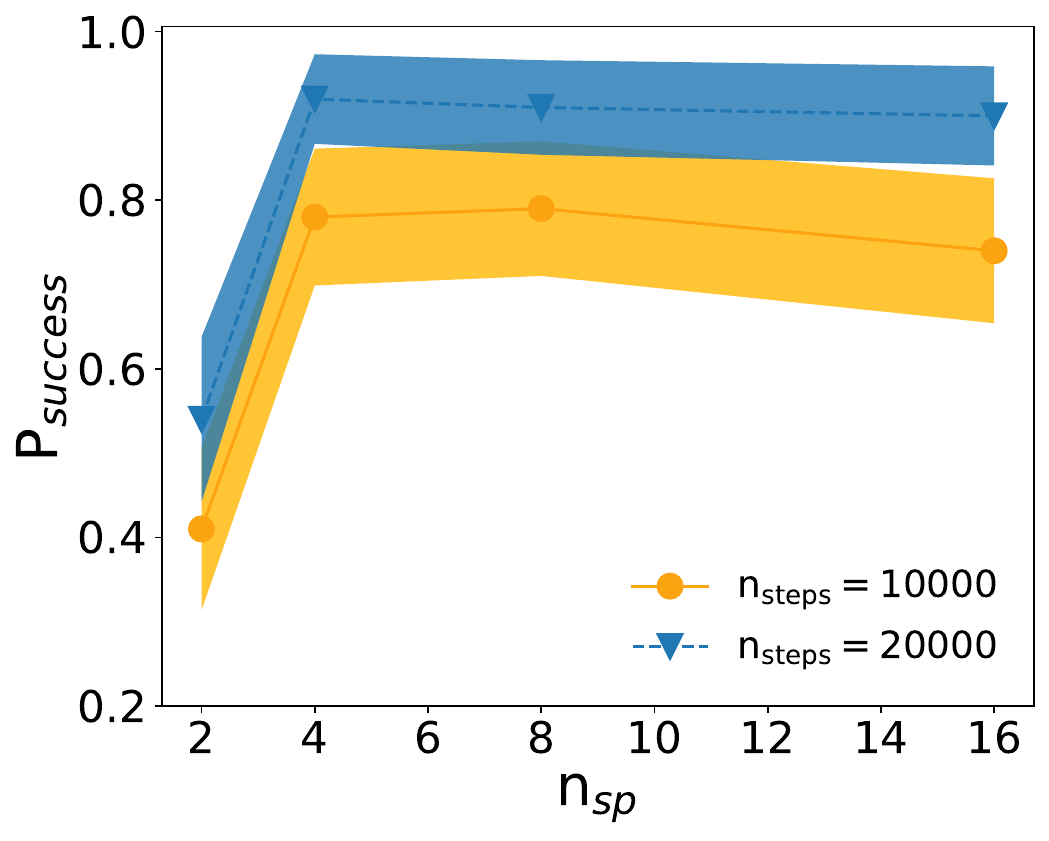}
\caption{\label{fig:Figs2} The probability of success, denoted as $P_{\text{success}}$, for training individual configurations with varying number of species $n_{sp}$ and a target of $\nu^*=-1$, is depicted for different maximum optimization steps. The coloured regions represent the $95\%$ confidence interval, indicating the estimated probability of success.}
\end{figure}

When training individual configurations, we observed that the probability of success $P_{success}$ improves as the number of optimization steps increases. As illustrated in Fig.~\ref{fig:Figs2}, $P_{success}$ increases by $10\%$ when employing twice the maximum optimization steps. This means that the $P_\mathrm{success}$ shown in Fig. 2 in the main text is only a lower bound on the actual probability of success. It is unclear if $P_\mathrm{success} \rightarrow 1$ if we were to wait long enough.

\subsection{Bond-level elastic response}

When a mechanically stable system is distorted by a symmetric strain tensor $\epsilon_{\alpha\beta}$, the change in energy can be written as
\begin{equation}
    \Delta E / V = \sigma_{\alpha\beta}^0 \epsilon_{\alpha\beta}+\frac{1}{2}\epsilon_{\alpha\beta}c_{\alpha\beta\gamma\delta}\epsilon_{\gamma\delta} + O(\epsilon^3)
\end{equation}
where $V$ is the volume of the system, $\sigma_{\alpha\beta}^0$ is the residual stress tensor and $c_{\alpha\beta\gamma\delta}$ is the elastic modulus tensor.

In general, there are 6 independent components of the elastic modulus tensor for 2d systems, which are calculated from the Hessian of the energy \cite{PhysRevE.90.022138}.

Because $\Delta E$ can be written as a sum over pairs of interactions, so too can the elastic modulus tensor,
\begin{equation}
    c_{\alpha\beta\gamma\delta}=\sum_{ij} c_{ij, \alpha\beta\gamma\delta}
\end{equation}
and the change in energy of a pair of interaction is $\Delta E_{ij}=\frac{1}{2}\epsilon_{\alpha\beta}c_{ij,\alpha\beta\gamma\delta}\epsilon_{\gamma\delta}$. With $c_{ij, \alpha\beta\gamma\delta}$ describing the ``bond-level" elastic response for a pair of interaction $ij$, it allows the calculation of the bond-level bulk modulus $K_{ij}$ and shear modulus $G_{ij}$.
 
\begin{figure}
\centering
\includegraphics[width=0.9\linewidth]{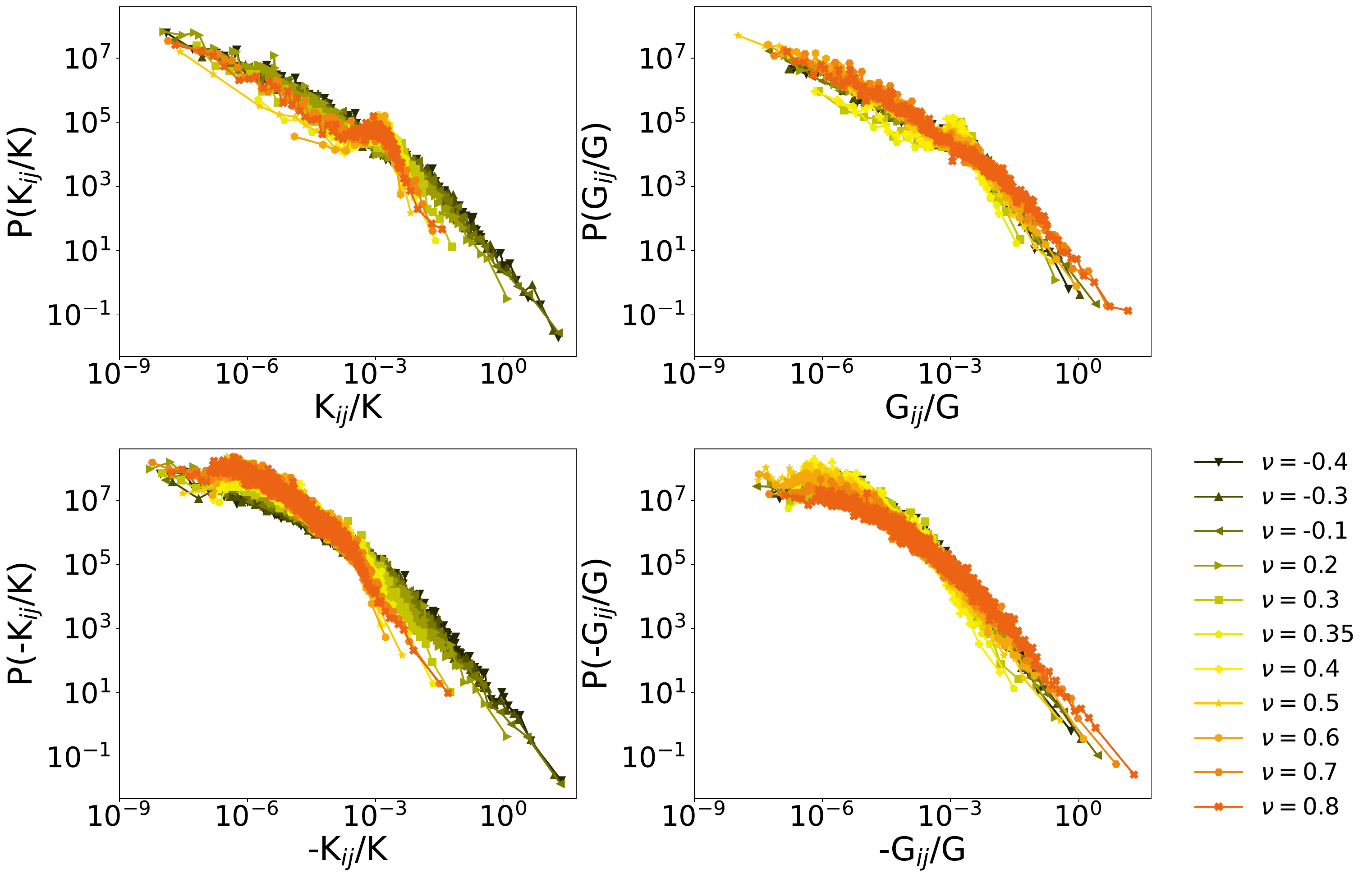}
\caption{\label{fig:Figs3} The probability distributions of the contribution of a pair of interactions to the bulk modulus $K_{ij}$ and shear modulus $G_{ij}$ for configurations with various Poisson’s ratios. These configurations were obtained by training configurations with two species. }
\end{figure}

\begin{figure}
\centering
\includegraphics[width=1.0\linewidth]{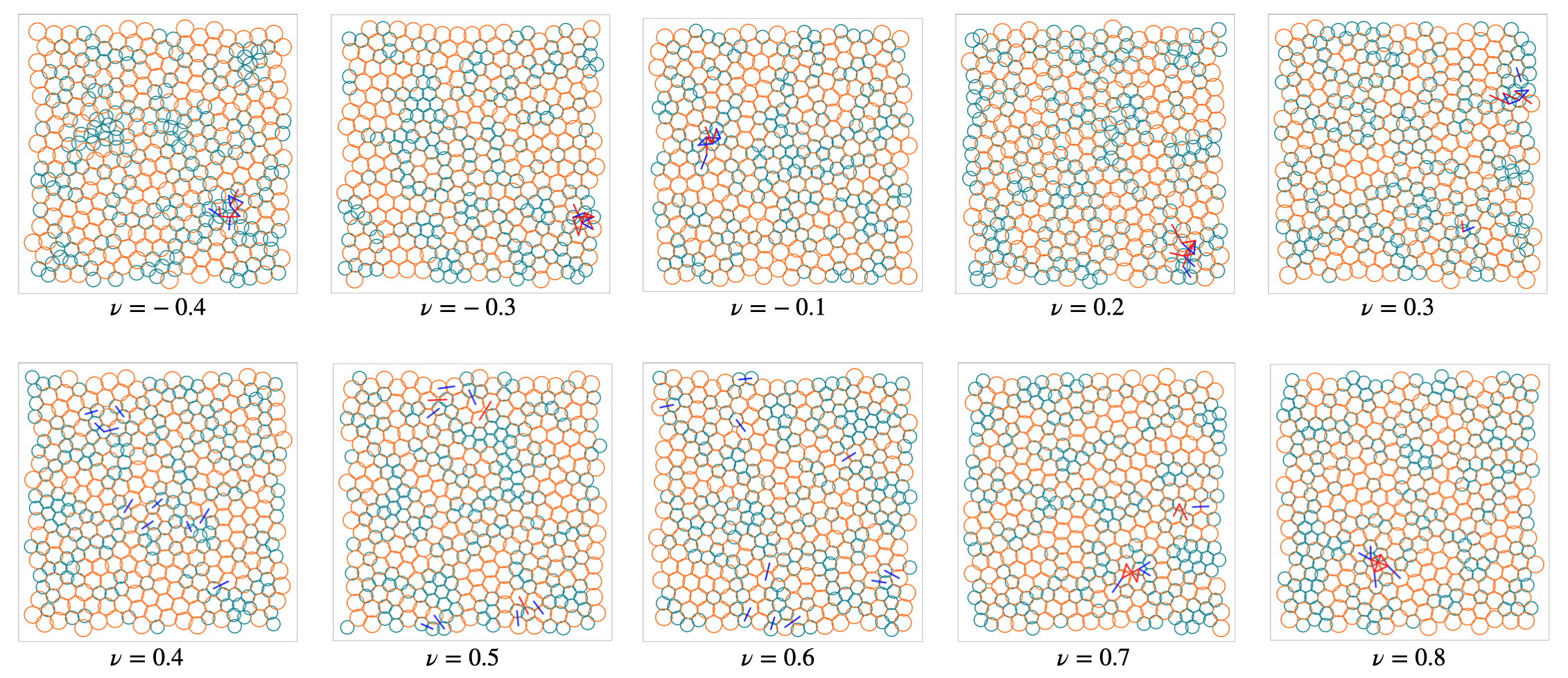}
\caption{\label{fig:Figs4} The packings with different Poisson's ratios display the 10 bonds with the largest absolute values of $K_{ij}$. These packings correspond to the examples presented in Fig~\ref{fig:Figs3}. Positive $K_{ij}$ are indicated by blue lines, while negative $K_{ij}$ are represented by red lines.}
\end{figure}

Figure~\ref{fig:Figs3} shows the probability distributions $P(K_{ij})$ and $P(G_{ij})$ at a range of Poisson's ratios $\nu$. As discussed in the main text, there is a noticeable change in the positive and negative tails of the distributions, especially $P(K_{ij})$, at different $\nu$. The emergence of these broad tails corresponds to the emergence of spatial localization in those largest values, as shown in Fig.~\ref{fig:Figs4}, meaning that the tails are spatially localized. This also roughly corresponds to when ensemble training of the Poisson's ratio becomes difficult.

\subsection{Ensemble optimization for desired Poisson's ratio $\nustar$}
\begin{figure}
\centering
\includegraphics[width=0.8\linewidth]{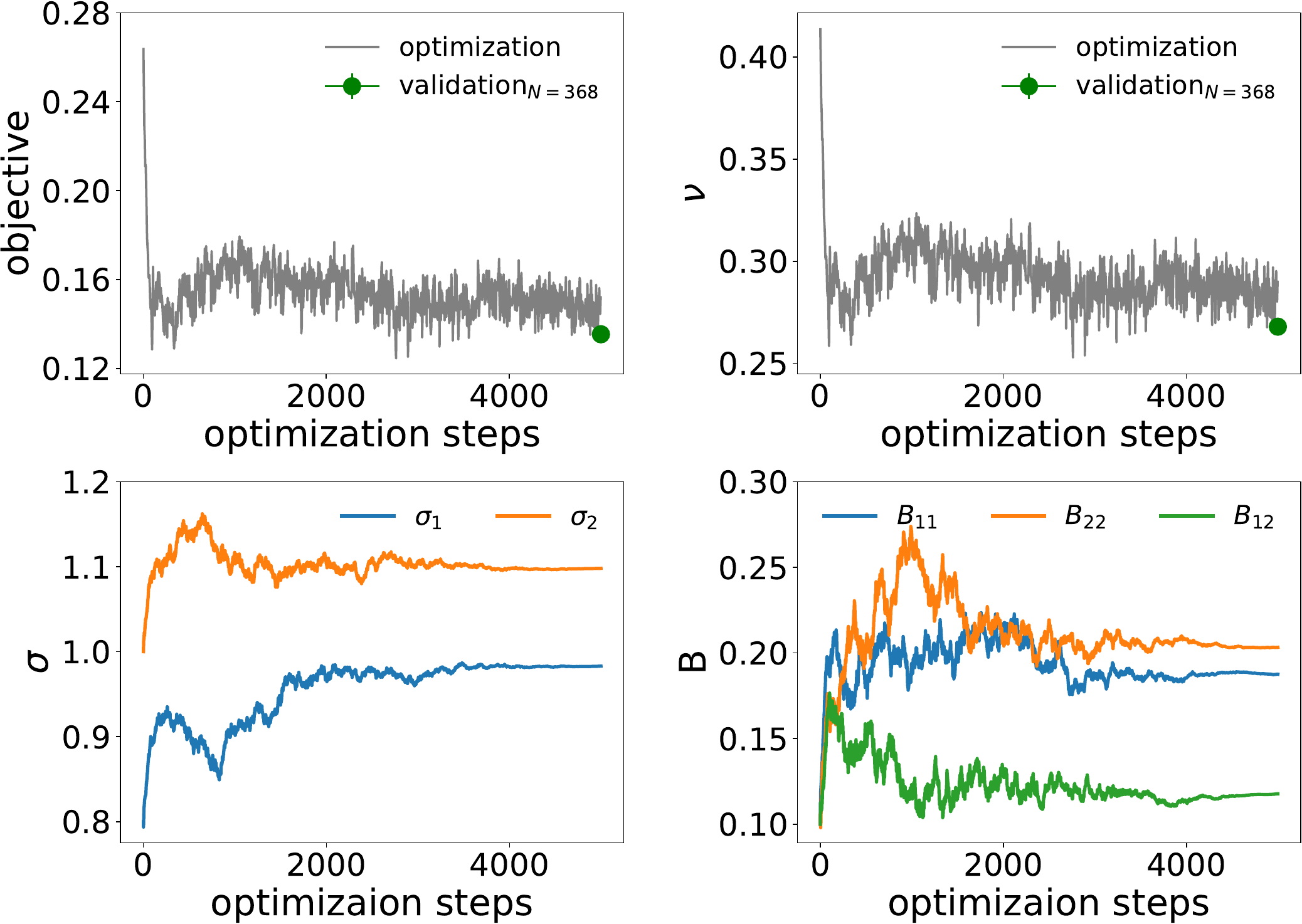}
\caption{\label{fig:Figs5} The optimization process for the ensemble with target $\nu^*=-0.1$. From top to bottom and left to right, the plots are the changes of objective function, $\nu$, diameters $\sigma_i$, and binding energies $B_{ij}$. }
\end{figure}

\begin{figure}
\centering
\includegraphics[width=0.8\linewidth]{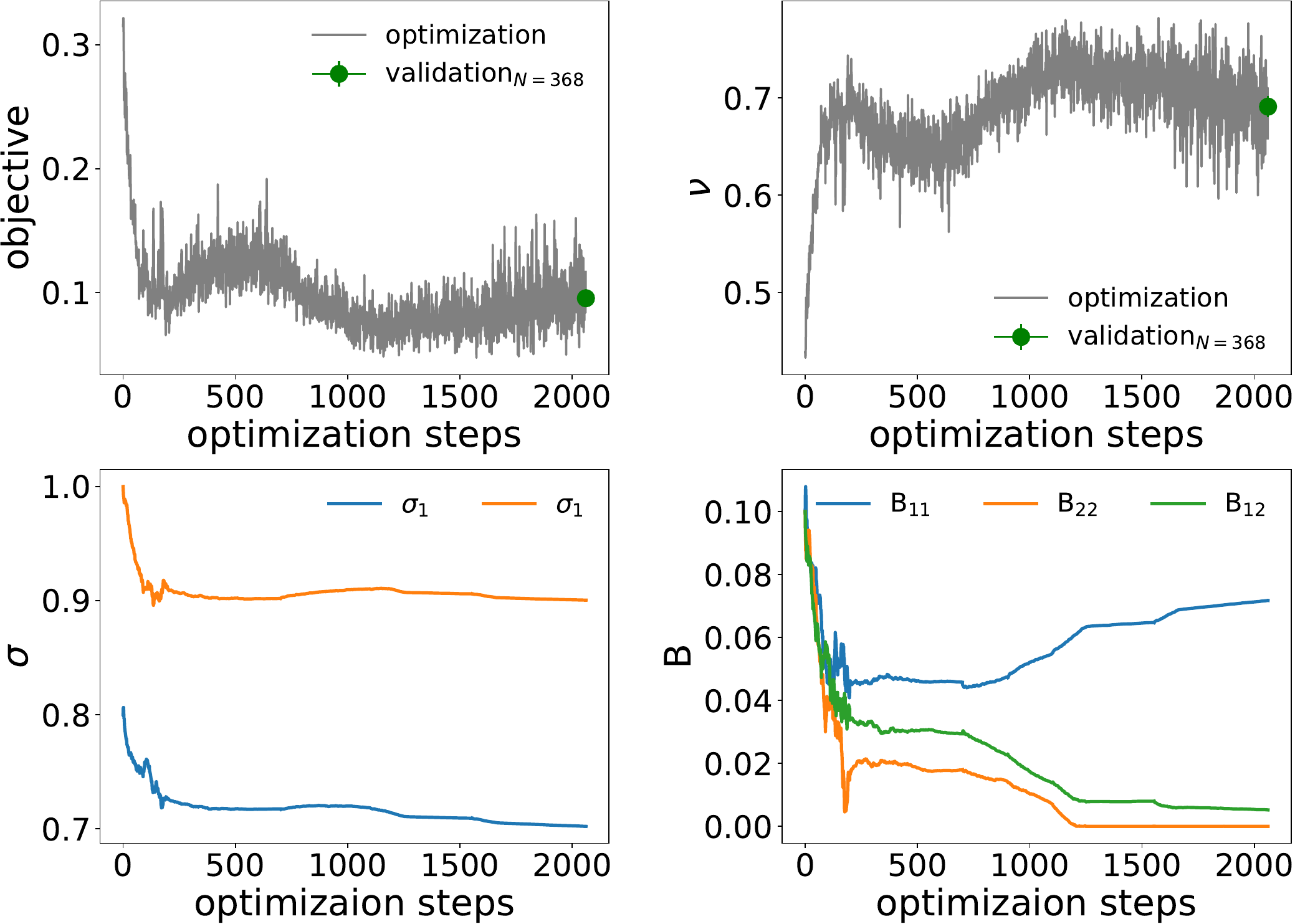}
\caption{\label{fig:Figs6} The optimization process for the ensemble with target $\nu^*=1.0$. From top to bottom and left to right, the plots are the changes of objective function, $\nu$, diameters $\sigma_i$, and binding energies $B_{ij}$. }
\end{figure}

In Fig.~\ref{fig:Figs5} and ~\ref{fig:Figs6}, we show the optimization processes for the ensemble with two extreme targets, $\nu^*=-0.1$ and $1.0$. The optimizations were performed for systems of $N=368$ particles evenly divided into 2 species. We use a variable learning rate $lr$ that changes for each step $s$ according to
\begin{equation}
lr(s) = \frac{lr_\mathrm{max}-lr_\mathrm{min}}{2}\left(\cos\left(\frac{\pi s}{s_\mathrm{max}}\right) + 1\right) + lr_\mathrm{min},
\end{equation}
where $lr_\mathrm{max}$, $lr_\mathrm{min}$, and $s_\mathrm{max}$ are manually adjusted for different training properties. For the results shown in Fig. 4 in the main text, we use  $lr_\mathrm{max} = 0.002$, $lr_\mathrm{min} = 10^{-5}$, and $s_\mathrm{max} = 5000$.

\begin{figure}
\centering
\includegraphics[width=0.5\linewidth]{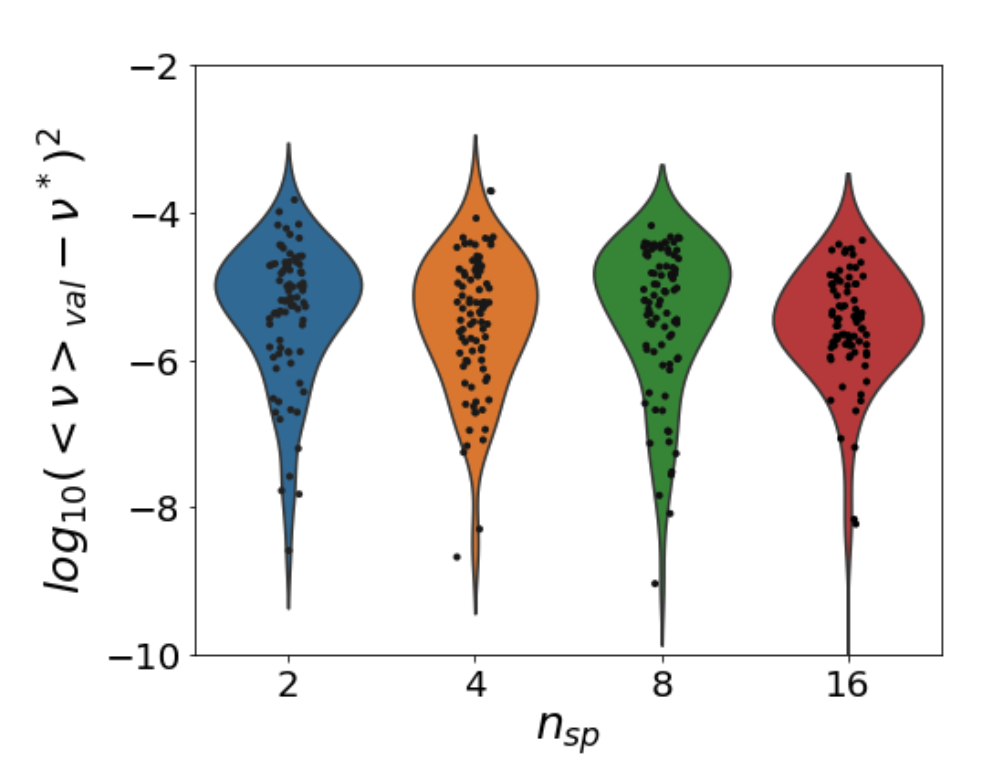}
\caption{\label{fig:Figs7} The violin plots of validations for the optimization for the ensemble with a target $\nustar=0.35$ across various species numbers. The black dots represent validations for 100 independent optimizations.  }
\end{figure}

Fig.~\ref{fig:Figs7} shows that the training process is reliable and reproducible. For a target of $\nustar=0.35$, we performed 100 independent optimizations each for $\nsp=2,4,8,16$. The optimizations were identical except for the seed for the random number generator that was used to choose the configurations to sample at each optimization step. As a result, the optimizations produced different solutions (sets of optimized parameters), but nevertheless resulted in very low validated losses, indicating that they were all successful.

\begin{figure}
\centering
\includegraphics[width=0.5\linewidth]{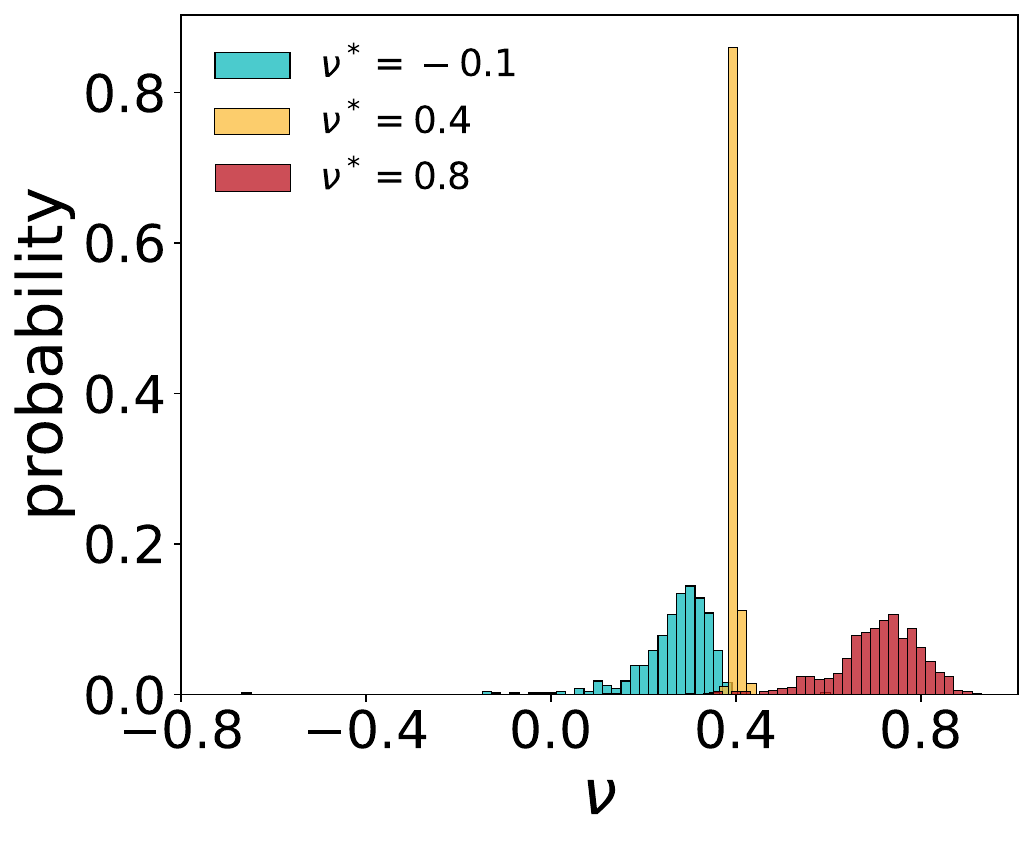}
\caption{\label{fig:Figs8} The probability distribution of validations for the ensemble optimizations with $\nu^*=-0.1,0.4,0.8$. }
\end{figure}

Fig.~\ref{fig:Figs8} shows the probability distribution $P(\nu)$ obtained during validation. For $\nu^*=0.4$, the distribution is narrow. However, for $\nustar=-0.1$ and $0.8$, the distributions become broader and exhibit long tails at small values, especially for negative Poisson's ratio.

\subsection{Ensemble optimization for desired pressure $p^*$}
\begin{figure}
\centering
\includegraphics[width=0.8\linewidth]{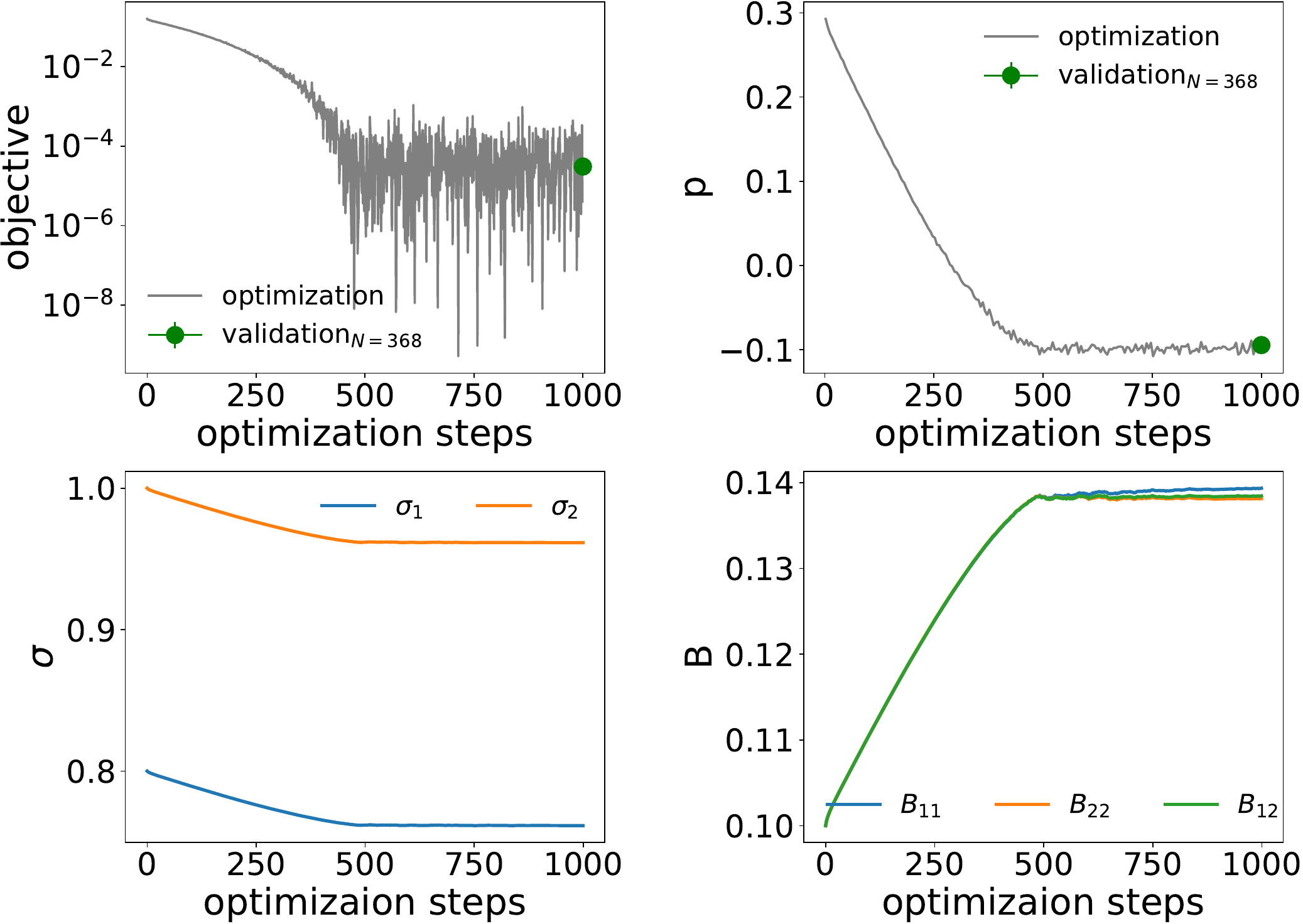}
\caption{\label{fig:Figs9} The evolution of an optimization process with target $p^*=-0.1$ for the ensemble. From top to bottom, left to right, the plots are the changes of objective function, pressure, diameters $\sigma_i$, and binding energies $B_{ij}$.}
\end{figure}

Our approach can be readily extended to address other system properties. Here, we have adapted the approach to control the pressure. As depicted in Fig.~\ref{fig:Figs9}, the magnitude of the objective function for targeting pressure is smaller than that for targeting Poisson's ratio. In addition, the parameter adjustments during optimization for pressure exhibit smoother transitions compared to those in the optimization for Poisson's ratio.

\subsection{Ensemble optimization for desired structural parameter $q_8^*$}
Although this paper primarily focuses on disordered solids, our approach can also be used to control, either by suppressing or promoting, measures of local structure. As an example, we consider the bond-orientational order parameter $q_8$, a measure of local 8-fold symmetry defined by $q_8=\frac{1}{N} \sum_{k=1}^N  |\frac{1}{n_b}\sum_{l=1}^{n_b} e^{8i\theta_{kl}}|$, where the inner sum is over the neighbours of $k$ and $n_b$ is the number of such neighbours. $q_8$ is a useful measure for detecting square lattices as it can include next-nearest-neighbor information.

\begin{figure}
\centering
\includegraphics[width=0.85\linewidth]{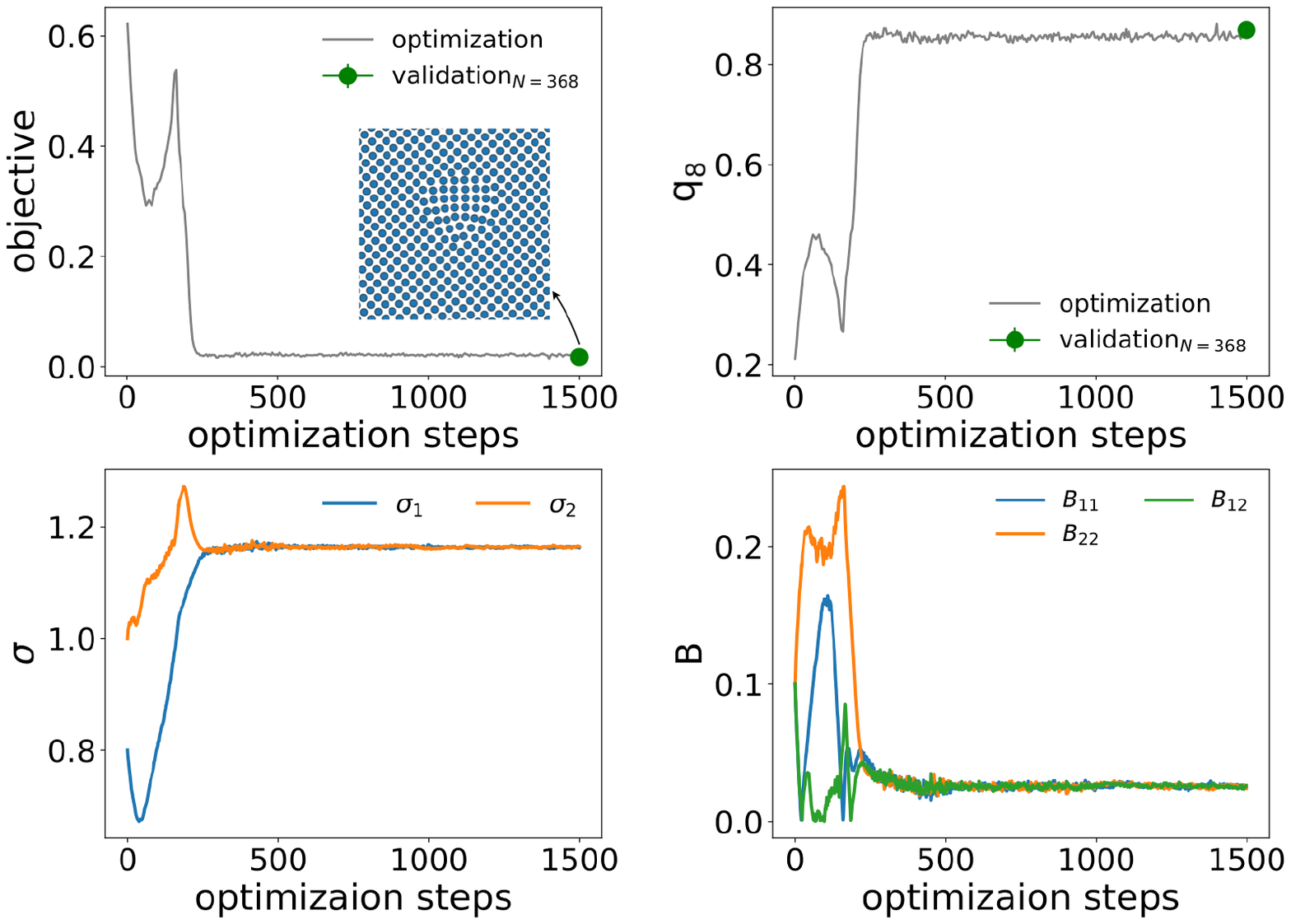}
\caption{\label{fig:Figs10} The evolution of an optimization process with target $q_8^*=1.0$ for the ensemble. From top to bottom, left to right, the plots are the changes of objective function, $q_8$, diameters $\sigma_i$, and binding energies $B_{ij}$.The insert shows an example configuration with parameters at the final optimization step, noting that the diameter of particles is shrunk by 0.5 times for clearer visualization.  }
\end{figure}

Figure~\ref{fig:Figs10} shows how we can learn to assemble a square lattice. This is a good example as it has a known solution of monodisperse repulsive particles at high densities\cite{10.1038/s41467-017-02316-3}.  By setting a target of $q_8^*=1.0$, we are able to recreate this solution in less than 500 optimization steps (note that the very low values of $B_{\alpha\beta}$ mean that the energy is dominated by the harmonic repulsive interactions). In this case, our objective does not quite vanish because our protocol of instantaneous cooling leads to the formation of polycrystals. As shown in the main text, we are able to precisely tune $q_8$ between roughly 0.2 and 0.9.

\subsection{Ensemble optimization for desired Poisson's ratio and pressure}
\begin{figure}
\centering
\includegraphics[width=0.7\linewidth]{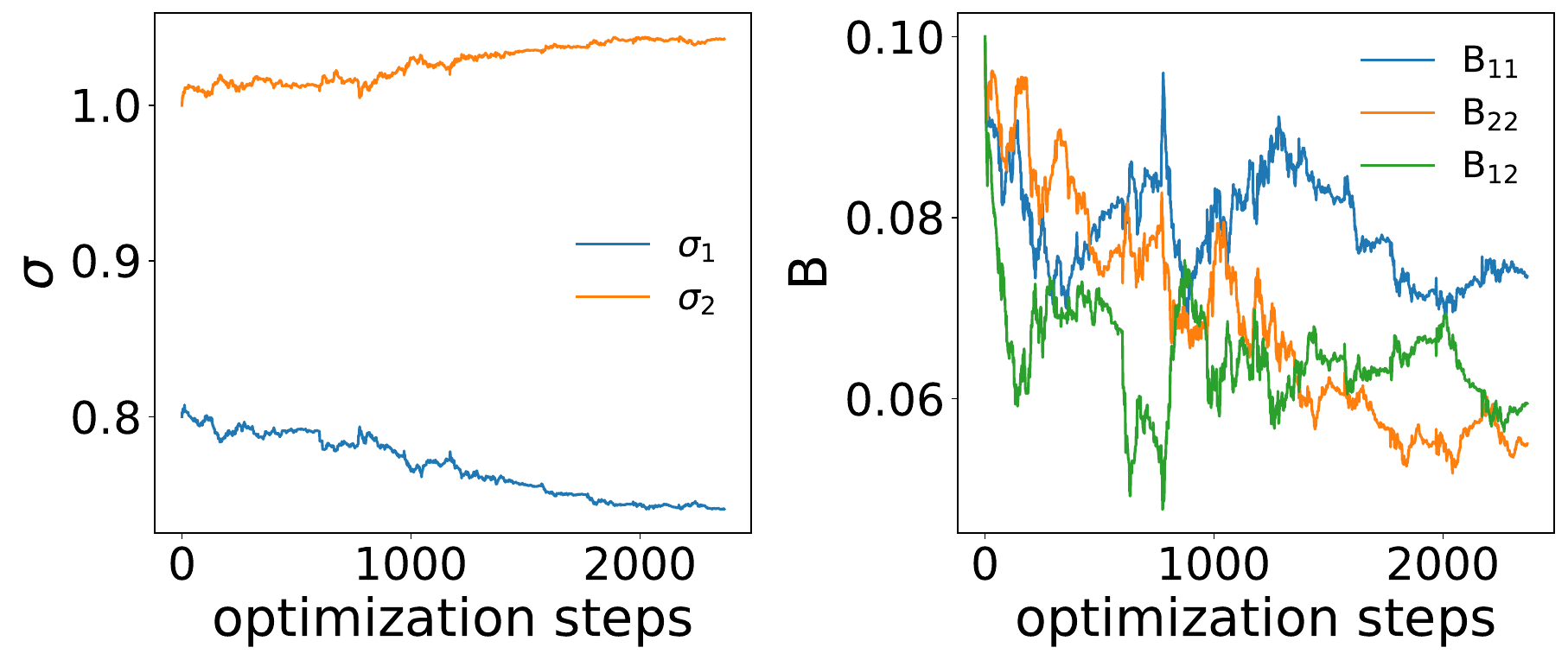}
\caption{\label{fig:Figs11} The evolution of parameters for the optimization with targets $\nu^*=0.5, p^*=0.4$ for the ensemble in the main text.  }
\end{figure}

Figure~\ref{fig:Figs11} illustrates the evolution of the design parameters for the optimization with targets $\nu^*=0.5$ and $p^*=0.4$ in the main text. 

\bibliography{supplementstuff}